\documentclass[aps,prd,twocolumn,amsmath,superscriptaddress,amssymb,showpacs,floatfix,nofootinbib,longbibliography,preprintnumbers]{revtex4-1}
\pdfoutput=1
\usepackage{graphicx}
\usepackage{bm}
\usepackage{times}
\usepackage[hidelinks]{hyperref}
\usepackage{slashed}
\usepackage{color}
\usepackage{aas_macros}

\usepackage{slashed}
\usepackage{lipsum}
\usepackage{subfigure}
\usepackage{multirow}
\usepackage{amsmath}

\usepackage{array} 
\usepackage{varwidth} 

\newcommand{\be}{\begin{equation}}
\newcommand{\ee}{\end{equation}}
\newcommand{\bea}{\begin{eqnarray}}
\newcommand{\eea}{\end{eqnarray}}
\newcommand{\epm}{e$^+$e$^-$}

\newcommand{\orcid}[1]
{\begingroup
\hypersetup{hidelinks}\href{https://orcid.org/#1}{\includegraphics[width=9pt]{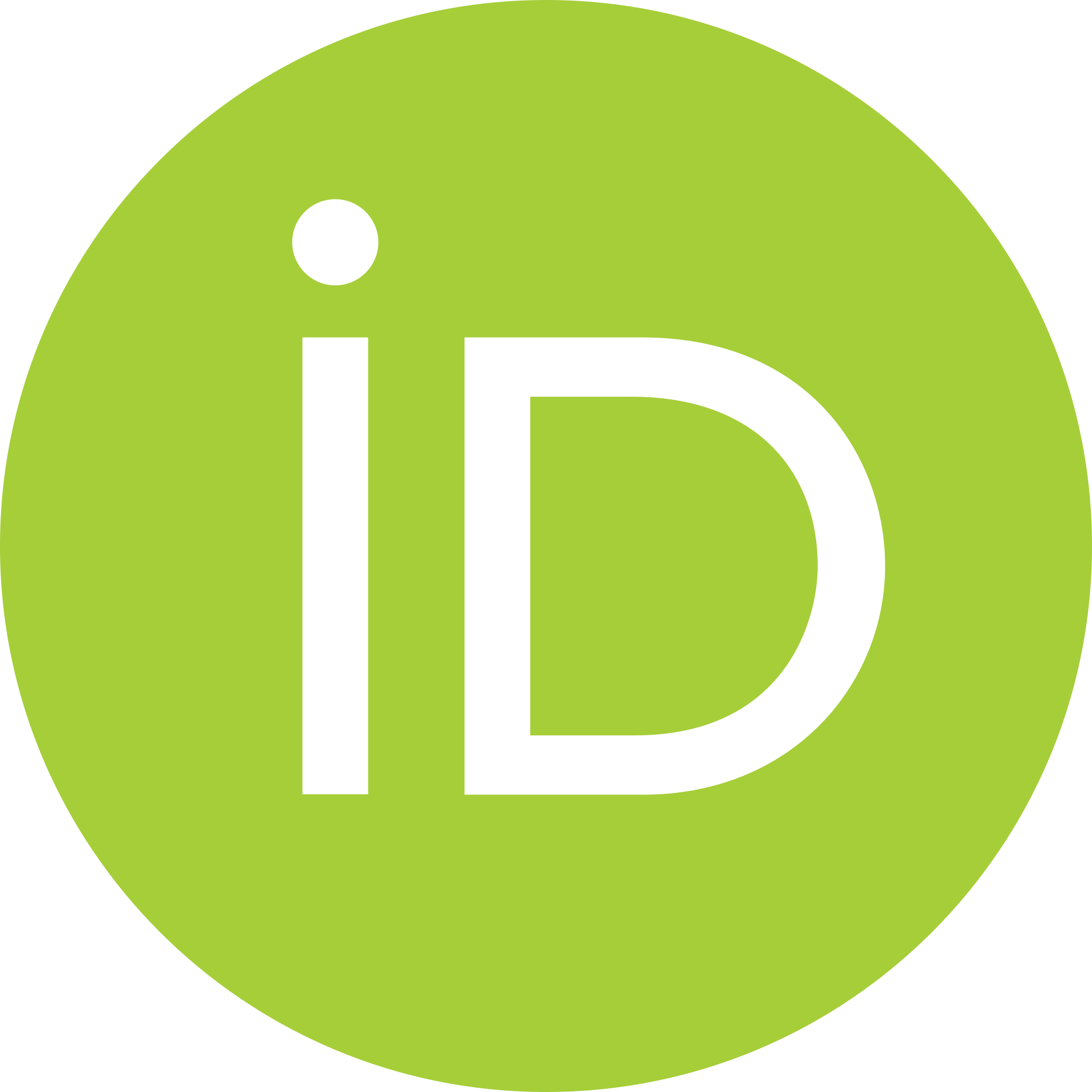}} 
\endgroup }

\begin{document}

\title{First Observations of Solar Halo Gamma Rays Over a Full Solar Cycle}
\author{Tim Linden
\orcid{0000-0001-9888-0971}}
\email{linden@fysik.su.se}
\affiliation{Stockholm University and The Oskar Klein Centre for Cosmoparticle Physics,  Alba Nova, 10691 Stockholm, Sweden}
\affiliation{Erlangen Centre for Astroparticle Physics (ECAP), Friedrich-Alexander-Universität \\ Erlangen-Nürnberg, Nikolaus-Fiebiger-Str. 2,
91058 Erlangen, Germany}

\author{Jung-Tsung Li \orcid{0000-0003-1671-3171}}
\affiliation{Center for Cosmology and AstroParticle Physics, The Ohio State University, Columbus, OH 43210, USA}

\author{Bei Zhou \orcid{0000-0003-1600-8835}}
\affiliation{Theory Division, Fermi National Accelerator Laboratory, Batavia, IL 60510, USA}
\affiliation{Kavli Institute for Cosmological Physics, University of Chicago, Chicago, IL 60637, USA}

\author{Isabelle John
\orcid{0000-0002-7604-1779}}
\affiliation{Dipartimento di Fisica, Universit\`a degli Studi di Torino, via P.\ Giuria, 1 10125 Torino, Italy}
\affiliation{INFN --- Istituto Nazionale di Fisica Nucleare, Sezione di Torino, via P.\ Giuria 1, 10125 Torino, Italy}

\author{Milena Crnogor\v{c}evi\'{c}~\orcid{0000-0002-7604-1779}}
\affiliation{Stockholm University and The Oskar Klein Centre for Cosmoparticle Physics,  Alba Nova, 10691 Stockholm, Sweden}

\author{Annika H.~G. Peter \orcid{0000-0002-8040-6785}}
\affiliation{Center for Cosmology and AstroParticle Physics, The Ohio State University, Columbus, OH 43210, USA}
\affiliation{Department of Physics, The Ohio State University, Columbus, OH 43210, USA}
\affiliation{Department of Astronomy, The Ohio State University, Columbus, OH 43210, USA}

\author{John F. Beacom \orcid{0000-0002-0005-2631}}
\affiliation{Center for Cosmology and AstroParticle Physics, The Ohio State University, Columbus, OH 43210, USA}
\affiliation{Department of Physics, The Ohio State University, Columbus, OH 43210, USA}
\affiliation{Department of Astronomy, The Ohio State University, Columbus, OH 43210, USA}

\preprint{FERMILAB-PUB-25-0301-T} 

\begin{abstract}
\noindent We analyze 15 years of {\it Fermi}-LAT data and produce a detailed model of the Sun's inverse-Compton scattering emission (solar halo), which is powered by interactions between ambient cosmic-ray electrons and positrons with sunlight. By developing a novel analysis method to analyze moving sources, we robustly detect the solar halo at energies between 31.6~MeV and 100~GeV, and angular extensions up to 45$^\circ$ from the Sun, providing new insight into spatial regions where there are no direct measurements of the galactic cosmic-ray flux. The large statistical significance of our signal allows us to sub-divide the data and provide the first $\gamma$-ray probes into the time variation and azimuthal asymmetry of the solar modulation potential, finding time-dependent changes in solar modulation both parallel and perpendicular to the ecliptic plane. Our results are consistent with (but with independent uncertainties from) local cosmic-ray measurements, unlocking new probes into astrophysical processes near the solar surface.

\end{abstract}

\maketitle


\vspace{-0.5cm}
\section{Introduction}

\begin{figure}[t!]
\centering
\includegraphics[width=.5\textwidth]{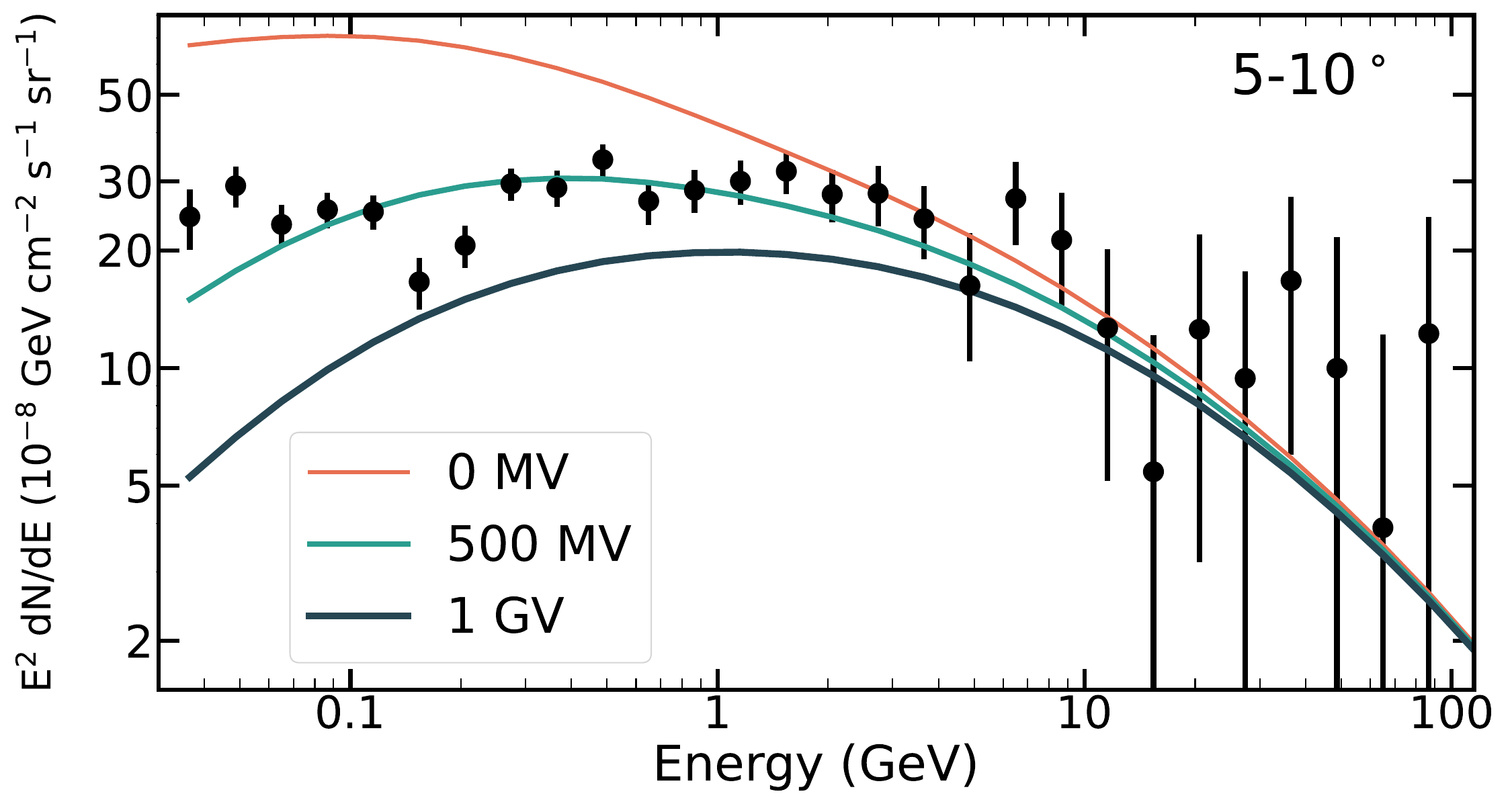}
\vspace{-0.2cm}
\includegraphics[width=.5\textwidth]{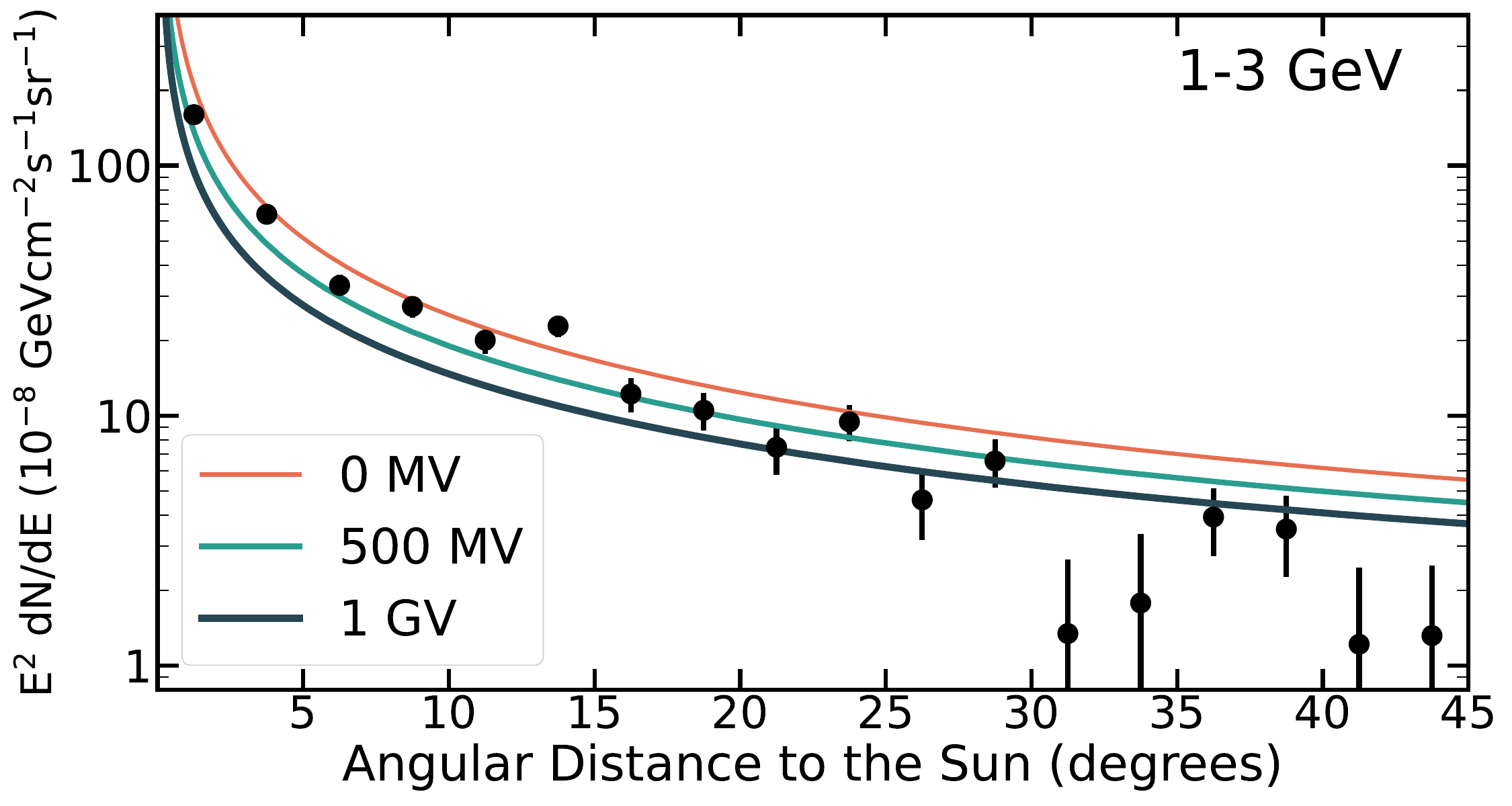}
\caption{Spectrum (top) and morphology (bottom) of the ICS halo surrounding the Sun in one representative angular bin (top) and one representative energy bin (bottom), compared to simple solar halo models that employ a force-field potential that is equivalent for both electrons and positrons and are \emph{not} fit to the data. Data are shown for our ``low-background map". The solar halo is robustly detected between 30 MeV and 100~GeV and out to 45$^\circ$ from the Sun. We find a close match between our models (colored lines) and the data.}
\label{fig:main}
\end{figure}

\noindent Despite being the preeminent object in our sky, the Sun is not an active high-energy $\gamma$-ray source. Its $\gamma$-rays reach only 4~GeV, and only during solar flares~\citep{Fermi-LAT:2011kqk, Fermi-LAT:2013vao, Fermi-LAT:2013cla, Pesce-Rollins:2015hpa, Omodei:2018uni, Share:2018kqt, Ajello:2021agj, Pesce-Rollins:2024kmy}. However, the Sun is an efficient \emph{passive} source that converts ambient galactic cosmic rays into $\gamma$-rays. There are two important processes. First, cosmic-ray protons and Helium nuclei can produce $\gamma$-rays via hadronic interactions with gas on the solar surface~\citep{Seckel:1991ffa, Zhou:2016ljf, 2020MNRAS.491.4852H, Mazziotta:2020uey, Gutierrez:2019fna, Gutierrez:2022mor, Li:2023twp, Puzzoni:2024enz, Li:2020gch, Ng:2024ksu, Griffith:2024jjy, BeckerTjus:2025zpa}. Second, cosmic-ray electrons and positrons (hereafter, \epm, with electrons being dominant) propagating through the heliosphere can inverse-Compton scatter (ICS) sunlight to $\gamma$-ray energies, producing a diffuse ``Solar halo"~\citep{Moskalenko:2006ta, Orlando:2006zs, Orlando:2008uk, Zhou:2016ljf, Orlando:2020ezh}.\footnote{Despite the fact that protons are much more abundant than electrons in the heliosphere, they produce negligible inverse-Compton emission because inverse-Compton scattering scales with $\gamma^2$, and the much heavier protons have relativistic velocities that are smaller by a factor of 1000.}

At high energies, the solar halo is easy to understand. The cosmic-ray \epm~density is roughly homogeneous and isotropic in the heliosphere. At very high energies ($>$10~TeV), the Larmor radius of these electrons is large compared to the solar system, and they essentially free-stream. These e$^+$e$^-$ upscatter solar photons that move radially outward from the Sun with a density that falls with the heliocentric distance as r$^{-2}$. Convolving these densities produces an r$^{-2}$ $\gamma$-ray morphology, which, when integrated over the line-of-sight, produces an azimuthally symmetric flux density that falls with the angular distance from the Sun as $\theta_s^{-1}$.

There are two complications. The first is that the ICS cross-section is proportional to the sine of the angle between the solar photon and the \epm. This, combined with the fact that conservation of momentum requires that the $\gamma$-ray travel in nearly the same direction as the \epm, significantly decreases the solar halo flux at angles near the Sun. The second complication occurs at low energies, where solar modulation induced by the heliospheric magnetic field decreases the \epm density, decreasing the low-energy ICS flux near the Sun.

While many $\gamma$-ray observations have probed the flux, spectrum, morphology, and variability of the solar disk across four decades in energy~\cite{Fermi-LAT:2011nwz, Ng:2015gya, Tang:2018wqp, Linden:2018exo, Linden:2020lvz, HAWC:2018rpf, HAWC:2022khj, Arsioli:2024scu}, there is only one published analysis of the solar halo. In 2011, Ref.~\cite{Fermi-LAT:2011nwz} used 1.5~years of Fermi-LAT data to detect the halo, finding that it extends between at least 0.1 to 10~GeV and out to at least 15$^\circ$ from the Sun. To perform this analysis, however, Ref.~\cite{Fermi-LAT:2011nwz} cut 93\% of the data to control the astrophysical backgrounds, and thus was unable to carefully study the temporal, spectral, and morphological properties of the halo.

In this paper, we significantly enhance our understanding of the solar halo by using 15 years of {\it Fermi}-LAT data and a radically new background model. We robustly detect the solar halo at energies between 31.6~MeV and 100~GeV, and out to 45$^\circ$ from the Sun, as shown in Fig.~\ref{fig:main}. Our precise measurements closely match theoretical models of ICS in the solar system, allowing us to provide novel measurements of solar modulation as a function of time, solar distance, and azimuthal angle. 

This paper is outlined as follows. In Sections~\ref{sec:data}~and~\ref{sec:background} we detail our {\it Fermi}-LAT analysis and our models for the solar halo. In Section~\ref{sec:results} we discuss the main results of our analysis, while in Section~\ref{sec:solarmodulation} we probe the effect of solar modulation, which is our most important theoretical uncertainty. Finally, in Section~\ref{sec:conclusions} we discuss the implications of our results for our understanding of cosmic-ray propagation in the heliosphere.


\section{Fermi-LAT Data Analysis}
\label{sec:data}

The analysis of solar $\gamma$-rays differs markedly from typical Fermi-LAT targets, due to the motion of the sun across the celestial coordinate system. In what follows, we detail the unique modeling codes which were tailored specifically for the analysis of moving sources. These codes have been made publicly available\footnote{https://github.com/trlinden/fermi\_solar\_halo}, along with documentation detailing their use.

We use exactly 15 years of {\it Fermi}-LAT data taken between 4 Aug 2008 and 4 Aug 2023. We analyze all \texttt{UltraClean} events (\texttt{evclass=512}) regardless of their PSFclass (\texttt{evtype=3}). We ignore events with a zenith angle exceeding 90$^\circ$ to remove Earth limb contamination. We bin our data into 28 logarithmic energy bins spanning between 31.6~MeV and 100~GeV. 

To accurately model a moving source, we use a two-step approach that converts both the {\it Fermi}-LAT data and exposure into helioprojective coordinates, which is a coordinate system centered on the Sun and defined such that $T_x$ is the angular coordinate that moves across the Sun's equatorial plane with positive values corresponding to the Sun's west limb, and $T_y$ is the coordinate that moves from the Sun's south pole towards its north pole. For events, we use the exact timing information of each $\gamma$-ray, along with its right ascension (RA) and declination (Dec), to determine its helioprojective position. 

Calculating the exposure in helioprojective coordinates is more complex, because the Sun moves quickly across the sky. Thus, we calculate the full sky exposure in increments of one hour, eventually producing $>$100,000 exposure cubes using the {\tt gtexpcube2} command from \texttt{Fermitools} (v. 2.1.30). Over each 1-hour period, the Sun moves approximately 360$^\circ$/365/24 $\sim$ 0.04$^\circ$, which is small compared to the $\sim$0.5$^\circ$ region over which {\it Fermi}-LAT exposures are typically binned. In each exposure, we convert between the Geocentric Celestial Reference System (GCRS) and helioprojective coordinate systems calculated at the median time of the 1-hour exposure. This allows us to (with significant computational resources) robustly calculate the solar exposure.

One complexity stems from the unique location of the Sun in {\it Fermi}-LAT instrumental coordinates. The Sun is almost always located within 3$^\circ$ of $\phi=0$, where $\phi$ is the azimuthal coordinate that wraps around the LAT boresight. This markedly differs from any other celestial source, which have relatively uniform distributions over $0^\circ \leq \phi \leq$ 360$^\circ$ over any lengthy time period. This unique distribution stems from the requirement that the {\it Fermi} solar panels, which are fixed at $\phi=0$, point towards the Sun. To account for this, we use ${\tt phibins = 60}$ to calculate the {\it Fermi}-LAT livetime cube, which accounts for the changing effective area as a function of $\phi$. Additionally, because many of these skewing events (which aim the solar panels at the Sun) occur quickly, we use the 1-s Fermi spacecraft ({\tt ft2}) files in our study. These alterations require significant computation, but affect our calculated exposure at the level of 1--2\%, which can have important effects.


\section{Background Modeling}
\label{sec:background}

After the exposure and $\gamma$-ray count maps are calculated in each 1-hour period, we produce a model for the $\gamma$-ray flux outside the solar system. We use a powerful data-driven $\gamma$-ray model first developed by our team in Ref.~\cite{Linden:2020lvz} to study emission from the solar disk, and subsequently Jupiter~\cite{Leane:2021tjj}. The key insight is that the {\it Fermi}-LAT has observed the entire sky during periods when the Sun is at other angular positions. This allows us to construct a background map of the $\gamma$-ray flux in each celestial coordinate without interference from the Sun. The predicted astrophysical background at the solar position is obtained by calculating the background flux at each point in RA and Dec, and predicting the number of astrophysical events at each point in $T_x$ and $T_y$ based on the distribution of RA and Dec exposures at each helioprojective coordinate. We calculate the number of expected counts in each T$_x$ and T$_y$ bin by summing over each one-hour exposure bin as follows:

\begin{equation}
    N_b(T_x, T_y) = \sum_{{\rm RA}, {\rm Dec}} \frac{N_{{\rm o}}({\rm RA}, {\rm Dec})}{\epsilon({\rm RA}, {\rm Dec)}} \times \sum_{{\rm t}} \epsilon^*({\rm RA}, {\rm Dec}, T_x, T_y, t) 
\end{equation}

\noindent where $N_b$ is the number of events expected from the astrophysical background in helioprojective coordinates, $N_{\rm o}$ is the total number of events at each point in RA and Dec calculated when the Sun was far away, $\epsilon$ is the total exposure of each celestial coordinate during this time, and $\epsilon^*$ is the specific exposure in each 4D slice that corresponds to a specific celestial and helioprojective coordinate at a time bin $t$. Note that the second summation occurs over each 1-hour time bin, while the outer occurs over the entire time period when the Sun is more than 60$^\circ$ away from a given position in RA and Dec.

To calculate N$_{{\rm o}}$ and $\epsilon$ without interference from the solar halo, we mask a 60$^\circ$ region around the solar position at each time. We also mask a 20$^\circ$ region around the Moon, due to the fact that the Moon is a bright low-energy $\gamma$-ray source and has a unique spatial correlation with the Sun~\cite{2012ApJ...758..140A}. Then, we use this model to predict the $\gamma$-ray counts map from background photons within our target region of interest (ROI), which consists of regions within 45$^\circ$ of the solar position.


\subsection{Additional Cuts}

In theory, this model would perfectly predict the $\gamma$-ray flux from sources that are not the Sun and Moon. It might include some emission from other solar system bodies not included in the analysis, but these have been found to be negligible~\cite{Leane:2021tjj, DeGaetano:2023suw}. In contrast to the techniques used in Ref.~\cite{Fermi-LAT:2011nwz}, this approach conserves nearly 100\% of the effective area in our model. 

However, we find that this model does not perfectly predict the background count map, and some residuals remain. Excitingly, these residuals are only 1--2\% of the diffuse normalization through most of our energy range, meaning that this data-driven diffuse model is \emph{significantly better} than any other diffuse model that has ever been used for {\it Fermi}-LAT studies, which often have a fidelity of $\sim$10--20\%. Unfortunately, because the diffuse $\gamma$-ray flux far from the Sun has a much higher surface brightness than the solar halo, this still significantly affects our results. Moreover, because the Sun is a unique object, occupying a singular position in $\phi$-space, it is impossible to produce blank sky studies to evaluate the performance of the diffuse model.

There are several potential sources of systematic error in our model. The first are variable sources, which may not have the same flux during periods when the source is inside or outside the ROI. Errors stemming from variable sources appear as bright ``streaks'' across the ROI, owing to their rapid motion from $+T_x$ towards $-T_x$ in the helioprojective frame. While this issue is certainly present, it is unlikely to significantly affect our results for three reasons: (1) the streaks from variable sources do not look like either halo or disk emission and should not be picked up by the solar templates, (2) there are just as many positive as negative residuals due to sources that are brighter inside or outside the ROI, and these tend to cancel, (3) the Sun has revolved through the celestial coordinate system 15 times in our analysis, which means that the effect from single outbursts will be reduced by the many rotations through the ROI. In our solar disk analysis~\cite{Linden:2020lvz}, these errors were entirely negligible, while in our Jupiter analysis~\cite{Leane:2021tjj} (which focused on a much dimmer source that only moved through celestial coordinates once), we found that several bright point sources near Jupiter needed to be removed to maintain the fidelity of our analysis at the very lowest energies ($\lesssim$100~MeV).

\begin{figure*}[t!]
\centering
\includegraphics[width=.99\textwidth]{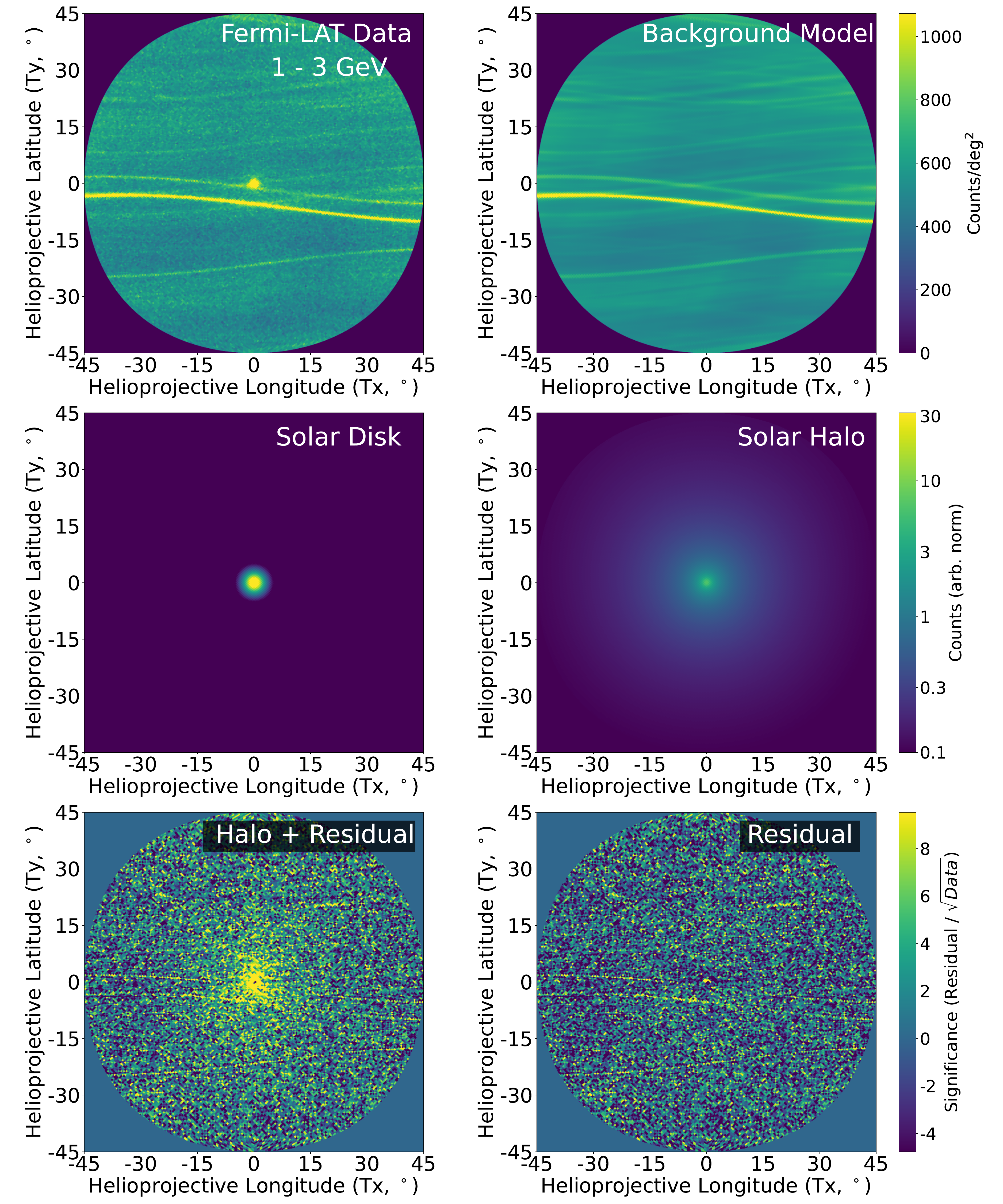}
\caption{The data, templates, and residuals in our analysis, evaluated at energies between 1 and 3.16~GeV in helioprojective coordinates. For this plot, we utilize the full dataset without point sources or the galactic plane masks ({\bf high-statistics map}), to convey the visual aspects of the plot more clearly. We show the {\it Fermi}-LAT data (top left) and the astrophysical background model (top right), noting the bright point sources that appear in each model. Arbitrarily normalized models for the solar disk (left) and solar halo (right) are shown on a log-scale in the second row, where we note the extension of the solar disk is driven by the $\sim$1$^\circ$ angular resolution in the 1--3.16~GeV bin.  The third row shows the significance of the remaining residual after the best-fitting background and solar disk model are subtracted, which represents the true solar halo emission along with any mismodeled residuals. The bright emission in the center is the solar ICS halo. The true residual after the best-fit solar halo model is also subtracted is shown in the bottom right.}
\label{fig:helioprojective_maps}
\end{figure*}

A second, potentially more nefarious, error may stem from $\phi$-dependent errors in the event-reconstruction. A similar error was uncovered by our group in Ref.~\cite{Tang:2018wqp}, which found that cosmic-ray backgrounds were slightly enhanced in the range 1--3~GeV in locations near the Sun compared to the rest of the sky. This systematic error was then verified by the {\it Fermi}-LAT Collaboration, and the cause of the excess was found to be cosmic-ray electrons that leak through the ribbons of the anti-coincidence detector more efficiently at angles normal to the LAT boresight (e.g., $\phi$=0$^\circ$, 90$^\circ$, 180$^\circ$, 270$^\circ$)~\cite{2018arXiv181011394B}. This issue was significantly diminished in the revised {\tt p305} dataset, but it is not clear that the leakage has been entirely removed, or whether there are additional $\phi$-dependent issues that could affect the analysis of the much-dimmer solar halo.~\footnote{Philippe Bruel, private communication (2023).}

Notably, both of these uncertainties scale with the background flux. Thus, regardless of the origin of the uncertainties, we can minimize their effect by cutting regions with high astrophysical backgrounds, at the cost of effective exposure\footnote{But seriously, if anybody can figure out why this happens, it would be awesome.}. Such a decision is warranted so long as the statistical uncertainties from the decreased exposure are smaller than the systematic issues relating to the diffuse analysis. In our default analysis we place the following additional cuts: (1) we remove all photons that are observed within 30$^\circ$ of the Galactic plane, or within 3$^\circ$ of a {\it Fermi}-LAT 4FGL source that has either a $\gamma$-ray flux above 2$\times$10$^{-9}$~ph~cm$^{-2}$~s$^{-1}$ in the energy range 1--100~GeV, or a variability index that exceeds 100. This includes 743 $\gamma$-ray sources. (2) We cut all time periods when the Moon lies within 60$^\circ$ of the Sun, effectively removing any emission from the Moon from our analysis. We note that, in the case of the solar disk and halo models, these masks must be integrated \emph{after} the solar models are smeared by the instrumental point-spread function, because while they represent complete masks in the celestial coordinate system, they are only partial masks in the heliocentric coordinate system. 

These cuts have the unfortunate effect of removing 81\% of the {\it Fermi}-LAT exposure and 92\% of the $\gamma$-rays from our study, but produce a map that has a significantly lower background rate. We call this the {\bf ``low-background map"}, and use this as the default analysis for our full 15~year time window, when the size of the statistical sample is sufficient. We show differences between this result and our {\bf ``high-statistics map"}, which does not include these cuts, in Appendix~\ref{sec:highstatistics}, and utilize results from our ``high-statistics map" when looking for yearly time variations, when a higher photon count is necessary.


\subsection{Models of the Solar Halo}
\label{subsec:solarhalo}

We construct theoretical models of the ICS $\gamma$-ray intensity in the solar halo following the approaches in Refs.~\citep{Moskalenko:2006ta, Orlando:2006zs, Orlando:2008uk, Zhou:2016ljf, Orlando:2020ezh}, specifically Eqs.~(1--7) in Ref.~\citep{Orlando:2008uk}. First, we integrate the ICS emissivity over the line-of-sight. In our calculations, we use the anisotropic Klein-Nishina cross section and treat solar photons as a blackbody, with a number density described in Eq.~(3) of Ref.~\citep{Orlando:2008uk}. We assume that the local angular distributions of cosmic-ray e$^+$ and e$^-$ are isotropic at any location in the heliosphere. The remaining quantity that needs to be modeled is the solar modulation effect of cosmic-ray e$^+$ and e$^-$ as they propagate toward the Sun. We note that these electrons and positrons can also produce synchrotron emission at lower wavelengths~\cite{Orlando:2022xsm}, which has not yet been observed.

We use the force-field approximation to describe solar modulation due to its simplicity and inclusion of energy losses~\citep{Gleeson:1968zza}. The force-field approximation is a steady-state diffusion-convection equation evaluated in the solar system frame. It assumes that cosmic-ray propagation is spherically symmetric with respect to the Sun  and excludes particle drift. Its solution connects the cosmic-ray e$^+$/e$^-$ intensity, $J_{\rm e^\pm}$, at the heliocentric distance $r$ to that in local interstellar space and is given by:
\begin{equation}\label{eq: force field potential}
    J_{\rm e^\pm}\left(r, E_{\rm e}\right) = J_{\rm e^\pm}\left(\infty, E_{\rm e}\right) \frac{E_{\rm e}^2 - E_0^2}{\left(E_{\rm e} + e \Phi\left(r\right)\right)^2 - E_0^2},
\end{equation}
where $J_{\rm e^\pm}\left(\infty, E_{\rm e}\right)$ is the local interstellar spectrum (LIS) of cosmic-ray e$^+$/e$^-$, $E_{\rm e}$ is the e$^+$/e$^-$ total energy, $E_0$ is the e$^+$/e$^-$ rest mass energy, $e$ is the elementary charge, and $\Phi \left(r\right)$ is the modulation potential at~$r$ (assuming zero modulation potential at infinity). For $J_{\rm e^\pm}\left(\infty, E_{\rm e}\right)$, we use the updated e$^+$ and e$^-$ LIS from Ref.~\citep{Bisschoff:2019lne}, which is computed using the GALPROP code with an empirical transport model fitting the observed cosmic-ray e$^+$ and e$^-$ spectra from Voyager~1 and PAMELA. For our first modulation potential model, $\Phi_1 \left(r\right)$, which we refer to as Model~I, we follow Eq.~(8) from Ref.~\citep{Moskalenko:2006ta}:
\begin{equation}
    \Phi_1 \left(r\right) = \Phi_0  \left(r^{-0.1} - r_b^{-0.1}\right) / \left({\left(1~{\rm AU}\right)^{-0.1} - r_b^{-0.1}}\right),
    \label{eq:model_original}
\end{equation}
where $\Phi_0$ represents the modulation potential evaluated at $1~{\rm AU}$ and $r_b$ is the radius of the heliospheric boundary, taken as $r_b=100~{\rm AU}$. This form of $\Phi_1 \left(r\right)$ is derived from the radial dependence of the particle mean-free-path during the Solar Cycle~21 minimum, using data from IMP~8, Voyager~2, and Pioneer~10~\citep{2005AdSpR..35..611F}.

We note that Model~I assumes an energy-independent form of $\Phi \left(r\right)$, which implicitly corresponds to an $1/f$ magnetic power spectrum over the entire frequency range (with $f$ being the magnetic fluctuation frequency). However, Ref.~\cite{Li:2022zio} shows that this assumption may lead to overmodulation at distances $r < 1~{\rm AU}$ by neglecting the inertial and kinetic ranges of magnetic turbulence. According to their theoretical model, the intensity reduction of cosmic-ray electrons and protons from $1~{\rm AU}$ to $0.1~{\rm AU}$ is estimated to be $\lesssim 10\%$. Observational data from Helios~1 and~2 show an even smaller reduction of only $2\pm 2.5\%$ from $1~{\rm AU}$ to $0.3~{\rm AU}$, although these measurements were specifically for cosmic-ray protons with kinetic energies between~$0.25$ and $0.7~{\rm GeV}$~\citep{2019A&A...625A.153M}. Similarly, data from the MESSENGER mission, collected between $0.3$ and $0.4~{\rm AU}$, suggest a radial gradient of less than $10\%$ per AU for cosmic-ray protons with kinetic energies above $125~{\rm MeV}$~\cite{2016JGRA..121.7398L}. To account for the potentially weak modulation for $r < 1~{\rm AU}$, we introduce a second modulation potential, $\Phi_2\left(r\right)$, referred to as Model~II:
\begin{equation}
    \Phi_2\left(r\right) = \left\{
    \begin{array}{ll}
        \Phi_0,                &  r < {\rm 1~AU}, \\
        \Phi_1 \left(r\right), &  r \ge {\rm 1~AU}.
    \end{array}
    \right.
    \label{eq:model_inner_no_modulation}
\end{equation}
In this model, there is no solar modulation for $r < {\rm 1~AU}$, thus providing an upper limit for the cosmic-ray \epm intensity within $1~{\rm AU}$ from the Sun, as also suggested in Ref.~\cite{Fermi-LAT:2011nwz}.

Last, to account for the energy dependence of solar modulation, we introduce a phenomenological model, referred to as Model~III, defined by
\begin{equation}
    \Phi_3 \left(r, E_{\rm e}\right) = \left\{
    \begin{array}{ll}
        \Phi_1 \left(r\right) \times \left( {E_{\rm e}}/{\rm 10~GeV} \right)^{-\alpha},    &  E_{\rm e} < {\rm 10~GeV},  \\
        \Phi_1 \left(r\right),    &  E_{\rm e} \geq {\rm 10~GeV}.
    \end{array}
    \right.
    \label{eq:model_energy_dep}
\end{equation}
where $\alpha \ge 0$. In this model, $\Phi_3 \left(r, E_{\rm e}\right)$ applies a stronger modulation to lower-energy cosmic-ray \epm and weakens at higher energies. By fitting both $\alpha$ and $\Phi_0$ to ICS data, this phenomenological approach has the advantage of revealing the general trend of cosmic-ray spectrum shapes without requiring detailed information about cosmic-ray transport.

In the three modulation models discussed above, we allow $\Phi_0$ to take different values for e$^+$ and e$^-$ to account for the charge dependence of solar modulation. These values are denoted as $\Phi_{0, {\rm e^+}}$ and $\Phi_{0, {\rm e^-}}$, respectively. Throughout the rest of this paper, we refer to Model~I as our ``default'' halo model, and refer to Models~II~and~III specifically, when they are employed. 


\subsection{Statistical Modeling}

Our background model automatically accounts for all point sources, extended sources, as well as the truly diffuse galactic, extragalactic and cosmic-ray backgrounds. Since this includes the vast majority of $\gamma$-ray emission, our statistical fitting procedure includes only a few components.

We fit the {\it Fermi}-LAT data within 45$^\circ$ of the Sun using a three-component model that includes: (1) our astrophysical background, (2) a solar disk model that employs a 0.26$^\circ$ radius source with a constant surface brightness and has a spectrum which is allowed to float independently in each energy bin, and (3) a model for the solar halo, described above. The solar disk and halo must first be convolved with the instrumental point spread function (PSF), which we calculate by averaging the point spread function at the location of the Sun in each time slice. The background model, on the other hand, is produced from real photons that have already been smeared by the instrument, and thus do not need to be reconvolved with a model PSF. We note that the 45$^\circ$ ROI in which we show our science results is smaller than the 60$^\circ$ ROI that we remove to create our background model (and utilize for the Moon cut). This ensures that there is limited leakage between the ``on'' and ``off'' regions of our analysis. 

We fit each component using {\tt iminuit}. In our default analysis, we fix the normalization of the background component to unity. This is warranted by the assumption that the background flux (from non-solar system sources) does not change based on the alignment of the solar system. It may be incorrect, however, if the effective area of the instrument has an unknown systematic error (not accounted for by {\tt gtltcube} with phibins=60), that changes the exposure based on the solar coordinate. 

The measured flux of the solar disk is not well-predicted by any theoretical models~\cite{Linden:2018exo, Tang:2018wqp, Linden:2020lvz}, and is allowed to float freely in each individual energy bin. At low energies, this component can be degenerate with the central part of the solar halo. However, at energies above a few GeV, changing the normalization of the solar disk has very little effect on our results.

The modeling choices that treat the solar halo itself are quite tricky. A minimalist model could be constructed under the following assumptions: (1) the background model is ``perfect", capturing all true astrophysical backgrounds that lie beyond the solar system to the level of Poisson noise, (2) the solar disk is well separated from the solar halo, leading to minimal degeneracies between the components, (3) there are no other important solar system objects (or they have been removed, e.g., by placing a cut around the Moon). In such a scenario, there is no need to model the solar halo at all --- its emission and spectrum are merely the residual that remains once the astrophysical background model is subtracted and the solar disk component is fit to the central source. In this case, a well-motivated choice would be to fit the background model (with a normalization of unity), and then fit the solar disk component, and call everything else the solar halo emission. This would be ``theoretically optimal" because our modeling would inject no prior knowledge about what the solar halo \emph{should} look like.

\begin{figure*}[t!]
\centering
\includegraphics[width=.99\textwidth]{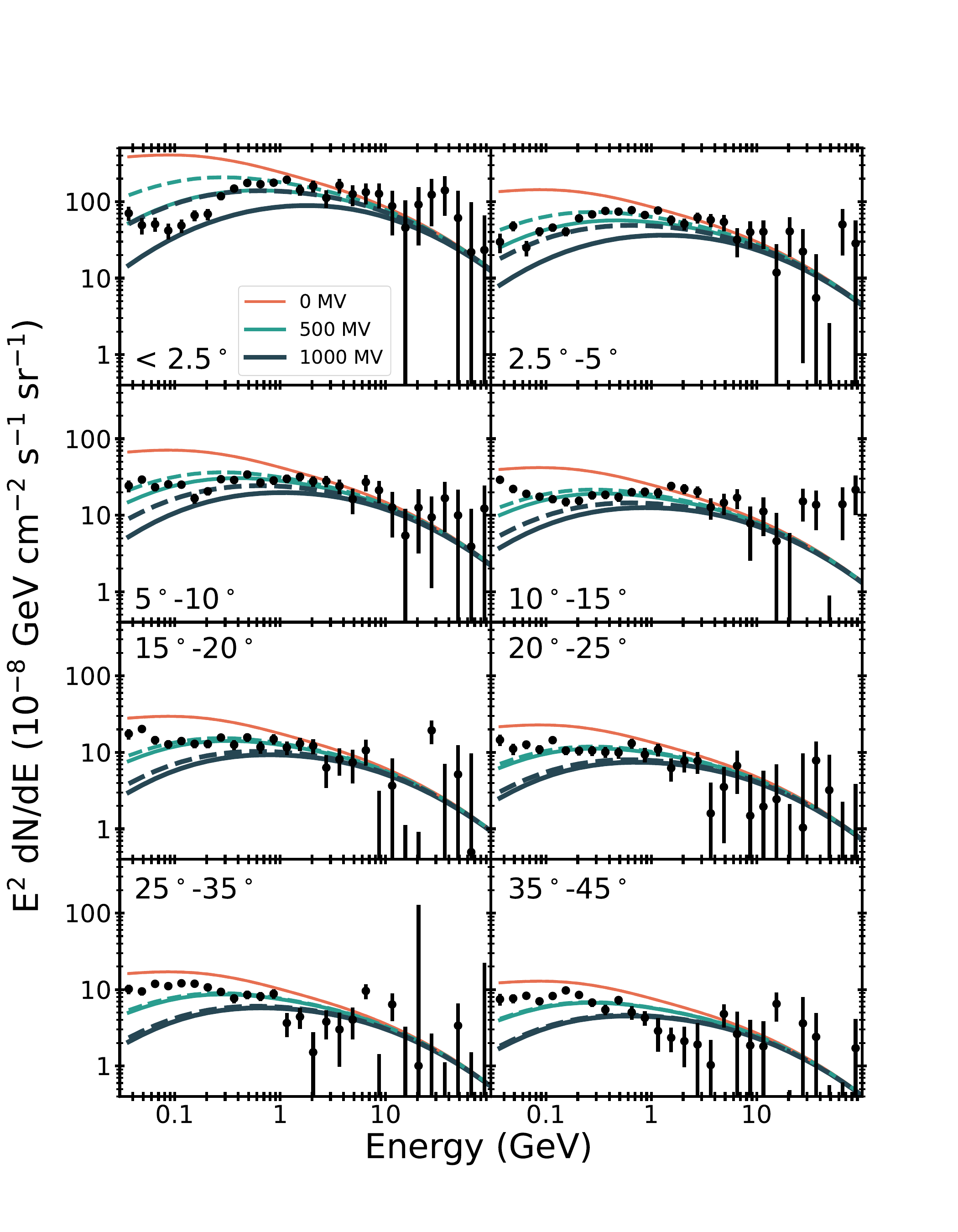}
\vspace{-1.1cm}
\caption{Same as Fig.~\ref{fig:main} (top), but showing the spectrum of the solar halo in eight different radial bins spanning between 0 and 45$^\circ$ from the Sun. The halo is robustly detected in the energy range from 31.6~MeV to 100~GeV. We show our theoretical models for ICS, including Model I (solid) and Model II (dashed). These models are not fit to the data, but instead follow a simple solar modulation potential that is fit to e$^+$e$^-$ data at the Earth position. This demonstrates the close match between theory and observation. We find that moderate modulation potentials near 500~MV provide the best fit to the $\gamma$-ray data.}
\label{fig:spectrum_in_all_radial_bins}
\end{figure*}

\begin{figure*}[p!]
\centering
\includegraphics[width=.99\textwidth]{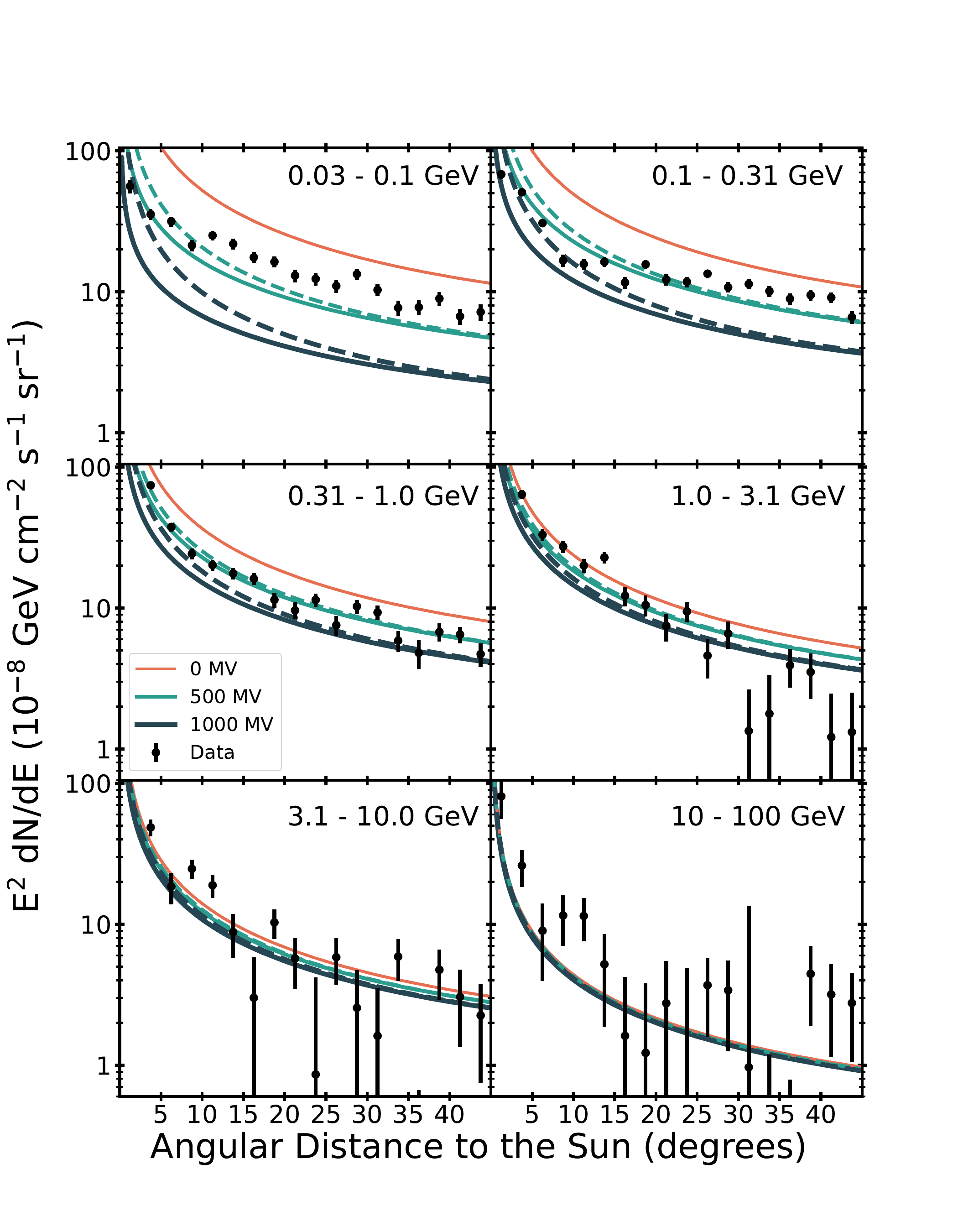}
\vspace{-1.5cm}
\caption{Same as Fig.~\ref{fig:main} (bottom), but showing the radial properties of the solar halo in six different energy bins spanning between 31.6~MeV and 100~GeV. The halo is robustly detected in every energy bin, and out to the 45$^\circ$ limits of our analysis. We show two solar modulation models, Model I (solid) and Model II (dashed), which are again not fit to the data, and thus demonstrate the close match between theory and observation. The preference for a modulation potential near 500~MV is primarily driven by low-energy data, which dominates the total $\gamma$-ray count rate. At high energies, we note significant systematic errors that exceed the statistical fluctuations of our analysis.}
\label{fig:morphology_in_all_energy_bins}
\end{figure*}

In this analysis, assumption (3) does hold~\cite{Leane:2021tjj, DeGaetano:2023suw}, assumption (1) should hold in principle, but appears to be violated in practice. However, assumption (2) is always violated at low energies, when the {\it Fermi}-LAT angular resolution is poor. Removing the solar halo from the model will cause the solar disk to become unphysically bright, as it tries to absorb nearby halo photons. This effect, discussed in Ref.~\cite{Linden:2020lvz}, requires us to input a solar halo into our fitting procedure to accurately calculate the solar disk flux. However, this creates a bias if the solar halo model is incorrect --- because the solar halo flux will be fit primarily by regions that are far from the Sun itself, and the input morphology of the solar halo would be forced to remove the specified number of photons from the inner regions based on the template morphology. 

In Ref.~\cite{Linden:2020lvz}, we accounted for this effect by binning the solar halo into multiple angular regions that floated independently. This works well, but at the cost of adding unphysical ``sawtooth" structures into the solar halo template. However, in this study (where we aim to study the solar halo, rather than just remove it) we instead adopt a two-component model, where the solar halo is the sum of two arbitrarily normalized templates with $\Phi_0$~=~1000~MV and $\Phi_0$~=~0~MV, which correspond to the maximal and minimal modulation potentials in this study.  The fit can change the normalization of these profiles in each energy bin to fit the inner and outer regions of the halo separately. At high energies, these profiles have similar morphologies, but the excellent {\it Fermi}-LAT angular resolution means that the disk and halo are no longer degenerate.

The final consideration is how to assign a flux to the solar halo. Corresponding to the above discussion, we could either: (1) report the flux picked up by the solar halo template(s), or (2) report the flux picked up by the solar halo template plus any remaining residual. The first choice, however, would be problematic, as we would be unable to discover any new solar halo effects that are not \emph{already} in our solar halo model --- we could only prove that the log-likelihood of solar halo templates provides a good fit to the data. This analysis would only allow us to examine the ``known unknowns" that we include in our model space, and would exclude the possible detection of e.g., changes in the radial profile of the halo emission or azimuthal asymmetries in the solar halo flux. Thus, throughout this paper, whenever we report the spectrum or morphology of the solar halo, we are reporting the total flux in our 45$^\circ$ after the diffuse background and solar disk emission are subtracted. When we study the best-fitting solar modulation parameters, we focus on the log-likelihood of these parameter models to our full dataset.

Figure~\ref{fig:helioprojective_maps} shows the helioprojective maps that correspond to each portion of this process. For this plot, we utilize the model of the {\it Fermi}-LAT data that does not have point source and galactic plane emission removed ({\bf the high-statistics map}), and choose an energy range of 1--3.16~GeV, which combines the fits from four independent energy bins. These choices are made to enhance the clarity of the plot and illuminate the underlying physics (instead of displaying spatial artifacts from our masking technique). Our results clearly demonstrate the close match between the background model and the {\it Fermi}-LAT data. Most notably, we clearly see that when the background model and an arbitrarily normalized solar disk model are subtracted, the existing residual (bottom left) shows clear indications of a bright solar halo. The morphology, spectrum, azimuthal distribution, and time variation of this signal is the topic of the rest of this paper.


\section{Results}
\label{sec:results}

Figure~\ref{fig:main} shows the main result of this paper, which includes both the spectrum and spatial extension of the solar ICS halo. In this plot, we focus on angles between 5 to  10$^\circ$ from the Sun and energies between 1 and 3~GeV, where the halo is robustly detected. We note that at this energy and angular range, the solar disk is well-separated from the solar halo, based on the $<$~1$^\circ$ angular resolution of the {\it Fermi}-LAT at GeV energies. This analysis uses our default model, which: (1) uses the {\bf low-background} map to decrease the issues with diffuse mismodeling, (2) fixes the normalization of the diffuse background to unity, (3) uses two solar halo models with modulation potentials of $\Phi_0=0$~MV and $\Phi_0=1000$~MV for both the electron and positron components (which float independently in each energy bin) to encompass uncertainties in the halo shape, (4) measures the resulting solar halo flux by reporting all photons that are not fit by the diffuse background or solar disk flux.

Excitingly, this simple model provides a close match between {\it Fermi}-LAT data and theoretically motivated solar halo models. In both cases, we obtain a solar halo flux of $\sim$3$\times$10$^{-7}$~GeV~cm$^{-2}$~s$^{-1}$ at energies between 1 and 3~GeV and distances between 5 and 10$^\circ$ from the Sun. We find that the $\gamma$-ray flux peaks at a few GeV and falls off at lower and higher energies. The morphology of the signal falls roughly as $\theta^{-1}$ between 5 and 30$^\circ$ from the Sun, before potentially falling off more rapidly at larger radii. These results extend the detection of the solar halo to an energy range between 31.6~MeV and 100~GeV and an angular range out to 45$^\circ$, significantly improving on previous work in Ref.~\cite{Fermi-LAT:2011nwz}. The solid lines show theoretically motivated fits for solar halo models based on Model I that have a normalization fit to the local observations of the e$^+$e$^-$ flux, and a radial and energy dependence based on the solar modulation models discussed in Section~\ref{subsec:solarhalo}. Because these are not fit to the data, they demonstrate the very close match between our theoretical model (Model I) and our observational results. Finally, we note that, despite the fact that we have removed about 90\% of the total $\gamma$-ray data, the statistical error bars (shown) throughout the majority of our analysis are extremely small and our uncertainties are dominated by systematics.

Figures~\ref{fig:spectrum_in_all_radial_bins} and~\ref{fig:morphology_in_all_energy_bins} extend the results of Figure~\ref{fig:main}, showing a range of bins for the energy and radial extent of the analysis, along with illustrative theoretical predictions that are based on simple solar modulation potentials with a normalization fit to the observed e$^+$e$^-$ flux near the Earth. Figure~\ref{fig:spectrum_in_all_radial_bins} shows that our models produce a good fit to the $\gamma$-ray data in regions spanning from $<$2.5$^\circ$ up to 45$^\circ$ from the Sun. We find that the normalization of the $\gamma$-ray emission is best fit by a small solar modulation potential of $\sim$500~MV when averaged over the full 15~year observation time of our study, which is in close agreement with modulation potentials calculated from observations of cosmic rays near Earth~\cite{1968ApJ...154.1011G, 2013LRSP...10....3P, Cholis:2015gna, Usoskin:2017cli, Corti:2018ycg, Wang:2019xtu, Kuhlen:2019hqb, Cholis:2020tpi}.

\begin{table}[]
    \centering
    \begin{tabular}{|c|c|c|c|}
    \hline
         $\Phi_{0, \rm e^-}$~(MV) & $\Phi_{0, \rm e^+}$~(MV)& d.o.f. & $\Delta$LG($\mathcal{L})$  \\
        \hline\hline
        0    & 0    & 0 & 867  \\
        100  & 100  & 0 & 2619 \\
        200  & 200  & 0 & 3375 \\
        300  & 300  & 0 & 3682 \\
        400  & 400  & 0 & 3767 \\
        475  & 0    & 2 & 3784 \\
        500  & 500  & 0 & 3724 \\
        600  & 600  & 0 & 3612 \\
        700  & 700  & 0 & 3461 \\
        800  & 800  & 0 & 3291 \\
        900  & 900  & 0 & 3104 \\
        1000 & 1000 & 0 & 2942 \\
        Four-Component & ---- & 108 & 3973 \\ 
        \hline
    \end{tabular}
    \caption{The improvement in the log-likelihood for models that include a theoretically motivated solar halo model (Model I) compared to a fixed background model that does not include the solar halo. There are no additional degrees of freedom in the halo fit, as the normalization is fixed by observations of the local electron and positron flux and the modulation potential of electrons and positrons at 1~AU are fit to specific values. The table provides representative values where the electron and positron modulation potentials are equivalent. However, we also provide the best fit value, which corresponds to a modulation of the electron flux by 475~MV and a modulation of the positron flux by 0~MV. The log-likelihood is reported for data taken between 31.6~MeV and 100~GeV, using our ``low-background" analysis, which removes approximately 90\% of the total $\gamma$-ray counts. We also provide the change in the log-likelihood for our default four-component model, which allows the normalization of a $\Phi_0=0$~MV profile and a $\Phi_0=1000$~MV profile to float independently for both electrons and positrons in each energy bin. This improves the log-likelihood by less than 200, despite including 108 additional free-parameters.}
    \label{tab:LGLs}
\end{table}

\begin{figure}[t]
\centering
\includegraphics[width=.5\textwidth]{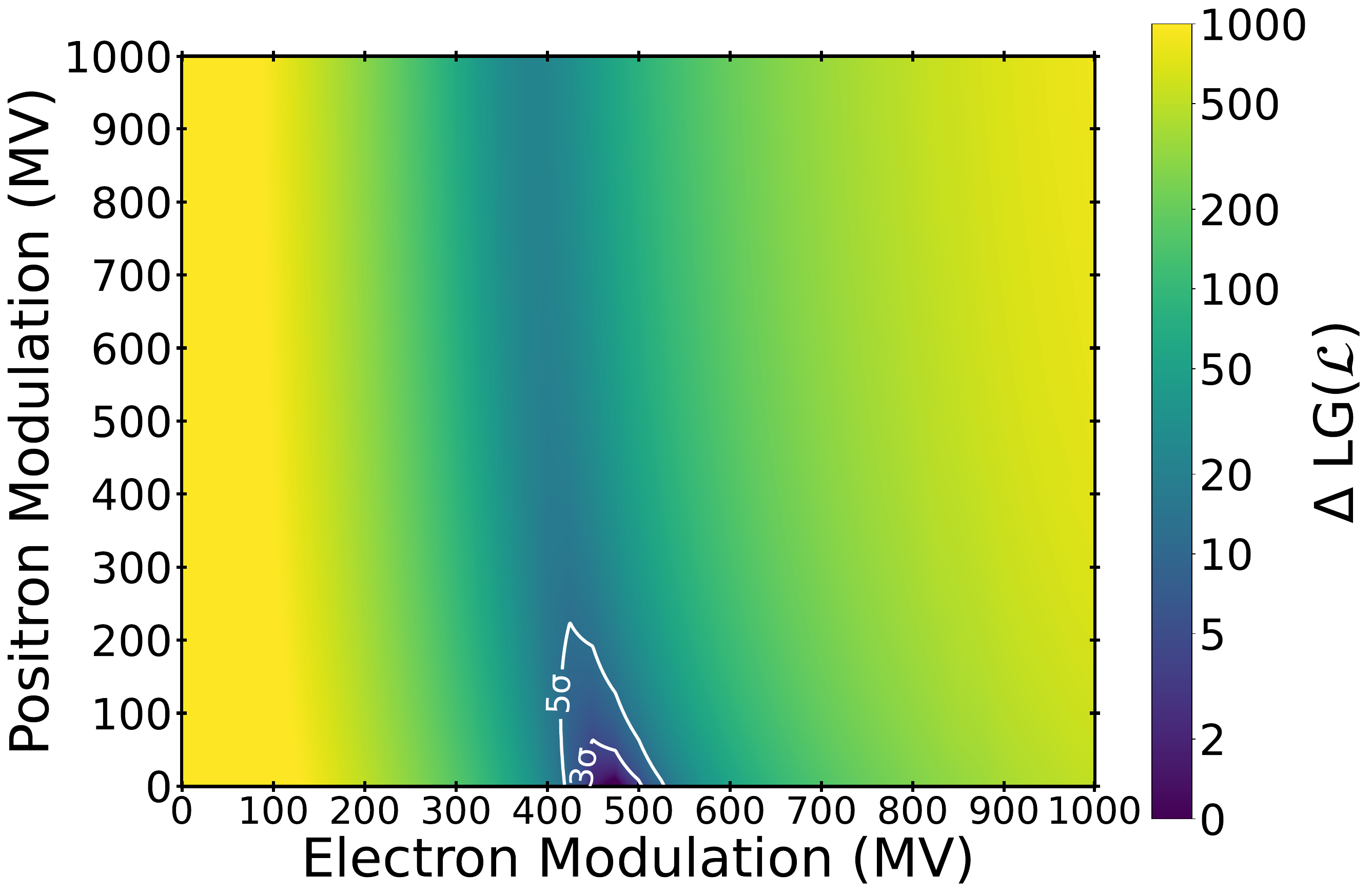}
\caption{The relative log-likelihood fit to the $\gamma$-ray data in our ROI as a function of the solar modulation potential defined at 1~AU for electrons (x-axis) and positrons (y-axis). The best-fit models have electron modulation potentials of $\sim$475~MV, with negligible modulation for positrons. The uncertainty on the dominant electron contribution is much smaller than the uncertainty on the smaller positron contribution.}
\label{fig:solarmodulation_heatmap}
\end{figure}

Figure~\ref{fig:morphology_in_all_energy_bins} shows what happens when we splice the same data in the opposite direction, showing the radial profile of the $\gamma$-ray data in six different energy ranges that span our analysis. This cut illustrates the radially dependent effect, and we again find that the best-fit values of the modulation at low energy move between moderate modulation potentials near the Sun, to relatively low modulation potentials far from the Sun. This deviation appears to point towards modulation effects beyond the simple potential we use in Model I. We note that the comparison between data and model (as shown in these plots) is not used to calculate the total log-likelihood at any point in our paper, but is shown only for visual comparison. All log-likelihood calculations are based on fits using a HEALpix map with nside=512 and eight logarithmic energy bins per decade. 

In Table~\ref{tab:LGLs}, we show the combined statistical result from these fits for our low-background cut, which affirms our interpretation of Figures~\ref{fig:spectrum_in_all_radial_bins}~and~\ref{fig:morphology_in_all_energy_bins}. We show the improvement in the log-likelihood (larger numbers are preferred) for models using different solar modulation potentials compared to the null hypothesis that no solar halo exists, which we set to a $\Delta$LG($\mathcal{L}$~=~0). We note that the $\gamma$-ray fluxes that correspond to these models correspond to the theoretical solar halo curves shown in Figures~\ref{fig:main}, ~\ref{fig:spectrum_in_all_radial_bins} and ~\ref{fig:morphology_in_all_energy_bins}, with spectra that are shown explicitly for the cases of 0~MV, 500~MV, and 1000~MV. We also show the results for our ``four-component" model, where the normalizations of four ICS components (based on solar modulation potentials of 0~MV and 1000~MV for both positrons and electrons) are allowed to float independently in each energy bin. We find two key results: (1) the simple theoretical models provide a very good fit to the main properties of the ICS halo, with log-likelihood values fairly close to the four-component fit despite having 108 fewer degrees of freedom, (2) the best-fit theoretical models have modulation potentials (which, in this case, are set to be equivalent for both positrons and electrons) on the order of 400~MV. We note that these are both ``averaged" modulation potentials that are calculated as single values over the full 15~year observation period of our study. 

Finally, Table~\ref{tab:LGLs} provides the best-fit log-likelihood for a two-dimensional fit of solar halo potentials for electrons and positrons, which is a theoretically well-motivated scenario due to the charge sign-dependent effects of solar modulation. In Figure Figure~\ref{fig:solarmodulation_heatmap} shows the two-dimensional fit of theoretically motivated solar halo models, based on a choice of different modulation potentials for positrons and electrons. Our best-fit model prefers a modulation potential of 475~MV for electrons, with no modulation (0~MV) of positrons. However, we note that the uncertainty on the positron modulation is quite large, due to the much smaller flux of the positron component. Calculating a one-dimensional uncertainty band for both electrons and positrons by minimizing over the other modulation potential provides a 5$\sigma$ uncertainty range of 474$^{+55}_{-59}$~MV for electrons and 0$^{+223}$~MV for positrons, where we do not let the modulation potential to fall below 0. 

We note that our observation that the positron modulation potential is much smaller than the electron modulation potential closely matches theoretical expectations for the positively charged Solar Cycle 24, and also matched observations of charge sign-dependent solar modulation of electrons and positrons near Earth~\cite{Cholis:2022rwf}. However, the quantitative value of the positron modulation potential in our analysis is likely unphysical due to systematic effects. Notably, because the $\gamma$-ray data cannot distinguish between ICS emission produced by an electron or positron, our fits only differentiate the components based on their input LIS spectra. The fact that positrons are highly sub-dominant (by a factor of $\sim$10-20) means that small errors in the electron LIS spectrum or modulation model will have outsized effects on positron modulation. Thus, while it is important to include contributions from both components, we only consider the electron modulation potentials to be well-modeled in our study, while, we consider the positron modulation value to primarily be a nuisance parameter that instead absorbs errors in our model.

Examining our results, and in particular the halo spectrum shown in Figure~\ref{fig:spectrum_in_all_radial_bins}, there is an indication of small (but only marginally statistically significant) anomalies that are worthy of comment. The first is at high energies, where we see a number of points that significantly exceed the predicted $\gamma$-ray flux. This includes a 30~GeV bump very close to the Sun, which is (on first inspection) interesting in light of the 30--50~GeV spectral dip observed from the solar disk in Ref.~\cite{Tang:2018wqp}. However, it would be difficult to correlate these two observations into a single coherent explanation, as the {\it Fermi}-LAT angular resolution at these energies is $\sim$0.1--0.2$^\circ$, which is far smaller than the scale of the ICS emission, and the emission processes between the disk and halo differ. We note that there are also high-energy ``excesses" in other spatial bins, but note that there is significant  point-to-point variation in the spectrum in this region. In summary, we do not believe that either of features --- although they may at first appear interesting -- are indications of any significant deviations from the theoretically predicted solar halo model.

The second spectral feature is more subtle, but critical for interpreting solar modulation models. We note that for low-energy data, higher solar modulation potentials ($\Phi_0 \sim$~500~MV) provide a better fit near the Sun ($\theta$~$\lesssim$~15$^\circ$). This change is statistically significant, due to the fact that the error bars in this region are small. We note that the standard 500~MV model is not a good fit to the data in the closest regions ($<$2.5$^\circ$) from the Sun. However, we note that this region encodes both the energy range and angular area where uncertainties in the Fermi-LAT exposure (which is changing rapidly at low-energies), and the point-spread function (which is largest at low energy) are worst. Similarly, we note that lower solar modulation potentials provide better fits farther from the Sun, which despite the lower statistical significance may be more important due to the lower smaller errors. It is important to note, that the total log-likelihood fit in our models is dominated by the high photon counts of relatively low-energy $\gamma$-rays. This tension motivates us to examine alterations to the standard force-field potential model (Model I) in the next section.

\section{A Closer Look at Solar Modulation}
\label{sec:solarmodulation}

The key science result of our paper is that the first-order properties of the solar halo are well fit by simple force-field solar modulation models within a $\gamma$-ray energy range spanning 31~MeV--100~GeV, and from the position of the Sun out to an angular distance of 45$^\circ$. Moreover, we note that the best-fit modulation potential for electrons calculated at 1~AU in ICS Model I ($\sim$475~MV) is consistent with the average modulation values calculated from fits to the local e$^+$e$^-$ data ($\sim$500~MV). However, our work in the previous section unveiled some regimes in which this simple model may not fit the $\gamma$-ray data. Thus, in this section, we closely probe alternatives to the simple force-field halo model (Model I).

\subsection{Changes in Modulation Potentials Near the Sun}
We note that the best-fit modulation potential for electrons calculated for the ICS halo using Model I ($\sim$475~MV) is very consistent with the average modulation values calculated from fits to the local e$^+$e$^-$ data (typically $\sim$500~MV). However, we note that these values need not be identical. First, the solar halo extends all the way to the Sun, measuring solar modulation in a regime that no cosmic-ray studies have probed. Second, the solar halo data probes the spherically symmetric modulation potential, which may differ significantly from the modulation potential in the ecliptic plane.

\begin{figure}[t]
\centering
\includegraphics[width=.5\textwidth]{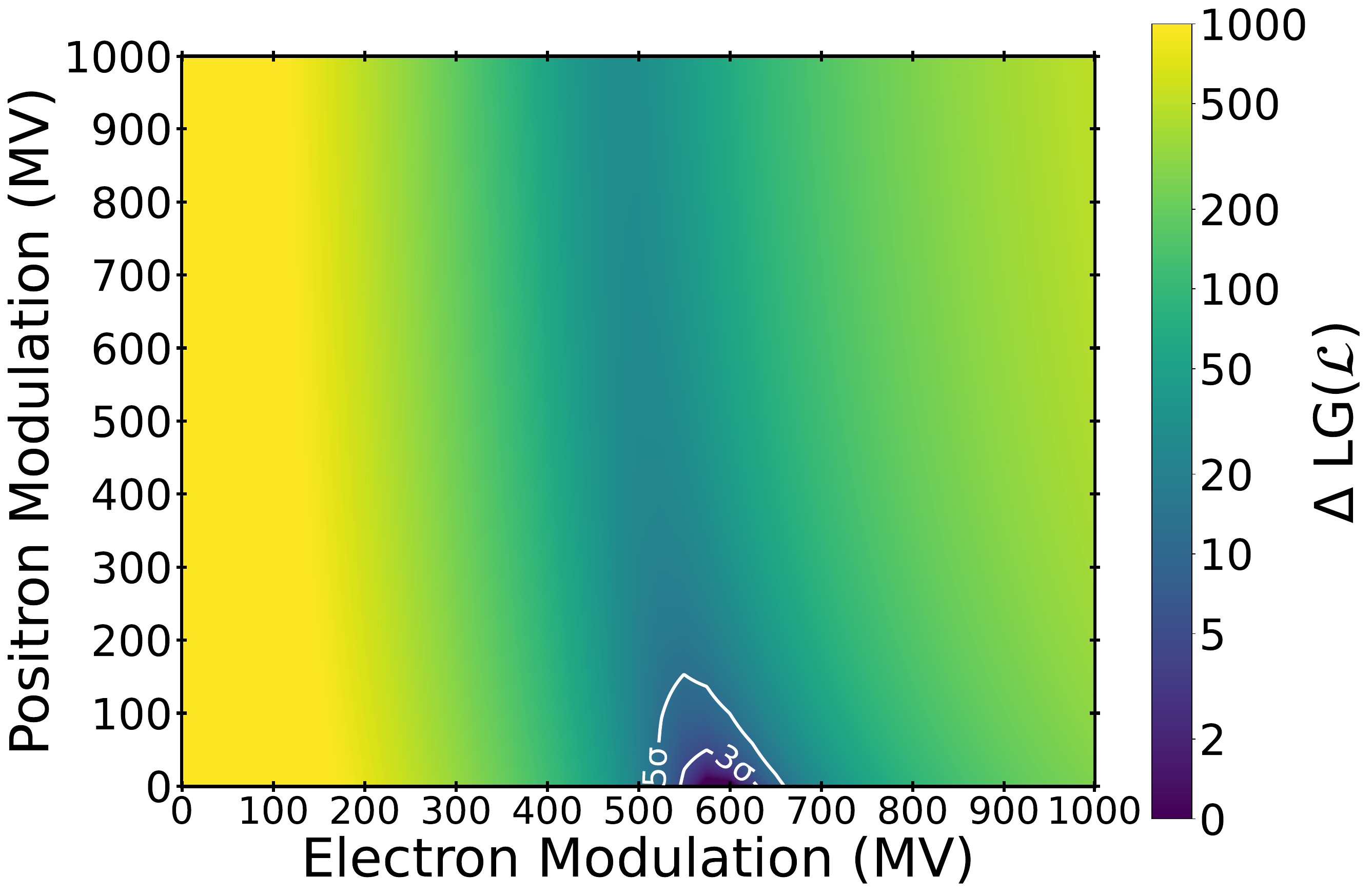}
\caption{Same as Figure~\ref{fig:solarmodulation_heatmap}, except for solar modulation Model~II, where there is no additional modulation between the Earth and Sun. The results are largely the same, but with a small increase in the solar modulation potential (to 600~MV) that offsets the lack of additional modulation between the Earth and Sun. }
\label{fig:solarmodulation_heatmap_jl}
\end{figure}

This leads us to consider a simple conceptual model of solar modulation that is split into two different regimes. The dominant effect of solar modulation takes place between the heliopause and the Earth, which lies outside our $\gamma$-ray ROI. In this regime, changing the modulation potential between 0~MV and 1000~MV decreases the overall normalization and hardens the spectrum of the solar halo, but does not significantly affect the $\gamma$-ray morphology determined by {\it Fermi}-LAT data.

The second regime corresponds to the subsequent solar modulation that occurs between the Earth and the Sun, and which lies inside our ROI. A key impetus for this paper is that solar modulation in this region has not been probed by cosmic-ray studies and is accessible only to our $\gamma$-ray observations. Modulation in this region controls the radial shape of the solar halo within our ROI, and also produces energy-dependent effects. However, the effect on the total normalization is smaller, as the majority of the modulation effect occurs between the edge of the heliosphere and 1~AU. Importantly, we note that the 45$^\circ$ edge of our ROI corresponds to electrons that have an average distance near 1~AU\footnote{At high energies, where solar modulation is not important, 50\% of the solar halo emission that is found 45$^\circ$ from the Sun originates from within 0.84~AU, while the other 50\% comes from emission at larger distances from the Sun.} from the Sun, meaning that the transition between these two regions occurs very close to the Earth's position. 

Interestingly, previous work in Ref.~\cite{Li:2022zio} used Parker Solar Probe data to study modulation between the Earth and Sun, finding weaker modulation within 1~AU compared to expectations from the force-field model. Motivated by their work, we consider an alternative model where there is no additional modulation between the Earth and Sun, leading to a total modulation potential given by Eq.~\eqref{eq:model_inner_no_modulation} in Model~II. In Fig.~\ref{fig:solarmodulation_heatmap_jl} we calculate the heatmap for the electron and positron modulation potentials in this model, finding the best-fit modulation potential for electrons increases to a value of 576$^{+84}_{-56}$ while the positron modulation potential remains consistent with 0$^{+153}$. This is consistent with the assumption that removing the additional effect of modulation near the Sun requires a larger modulation potential in regions outside 1~AU.

By comparing the log-likelihoods of our best-fit models, we find that halo Model~I (which does include solar modulation between the Earth and Sun) fits the data somewhat better than Model II, with a modest log-likelihood preference of 26. We note that we do not consider this a statistically significant exclusion of Model II, due both to the fact that our uncertainties are systematics dominated, and that the change in log-likelihoods is similar to the changes in log-likelihoods stemming from the 25~MV binning of our modulation potentials. However, we also find no evidence supporting such an alteration to the standard force-field model. This motivates us to instead look at alternative scenarios for changing the solar modulation potential.


\subsection{Energy Dependence of Solar Modulation}

In order to better understand the effect of solar modulation on the ICS halo, we now divide our analysis into different energy bins and calculate the best-fit modulation potential in each. We stress that these energy bins correspond to the $\gamma$-ray energies in our analysis, and not the \epm~energies. This means that each $\gamma$-ray energy bin corresponds to a range of \epm~energies based on the \epm~spectrum and the ICS kernel that relates the electron and $\gamma$-ray energies. 

\begin{figure}[t]
\centering
\includegraphics[width=.5\textwidth]{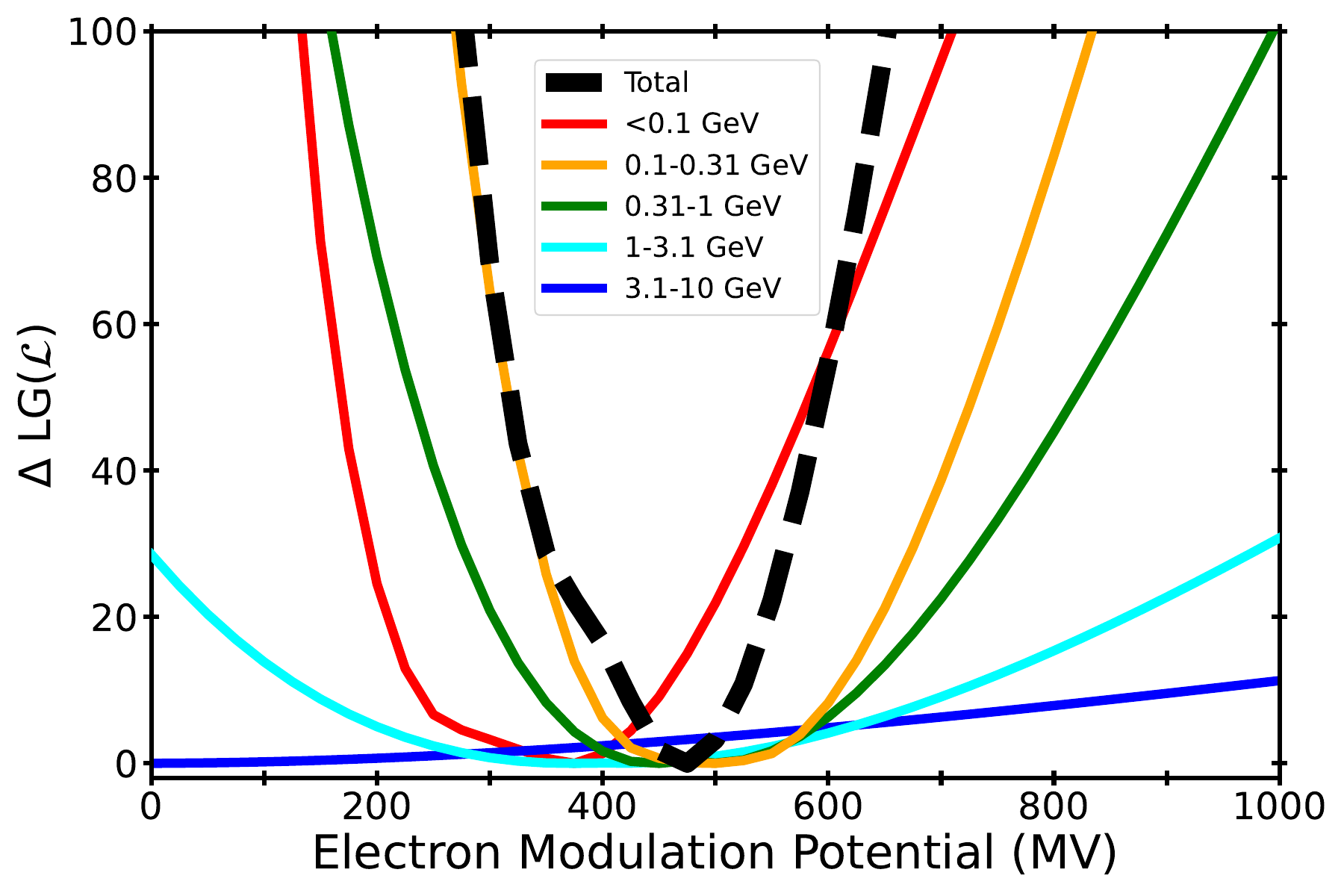}
\caption{The fit of different solar modulation potentials for electrons within different $\gamma$-ray energy ranges using solar modulation Model I. Each range is the sum of four energy bins in our analysis (except for the highest bin, which sums the contributions from eight bins). We find a broadly consistent preference for modulation values between 400 and 600~MV in all energy bins, though with a slight preference for negligible modulation in the highest $\gamma$-ray energy bins (which correspond to energy ranges where the \epm modulation would be negligible).}
\label{fig:solarmodulation_energybins}
\end{figure}

\begin{figure*}[t!]
\centering
\includegraphics[width=.48\textwidth]{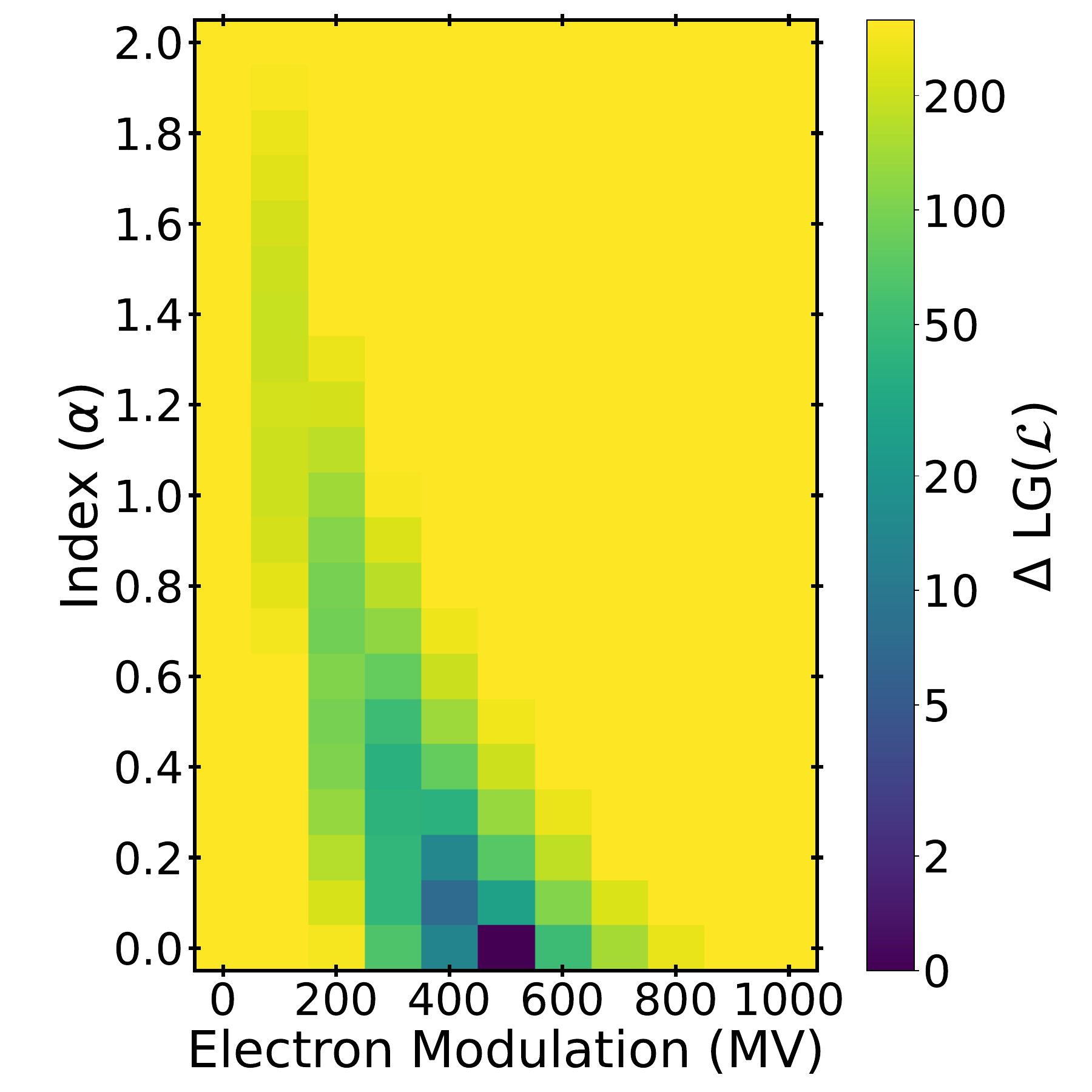}
\includegraphics[width=.48\textwidth]{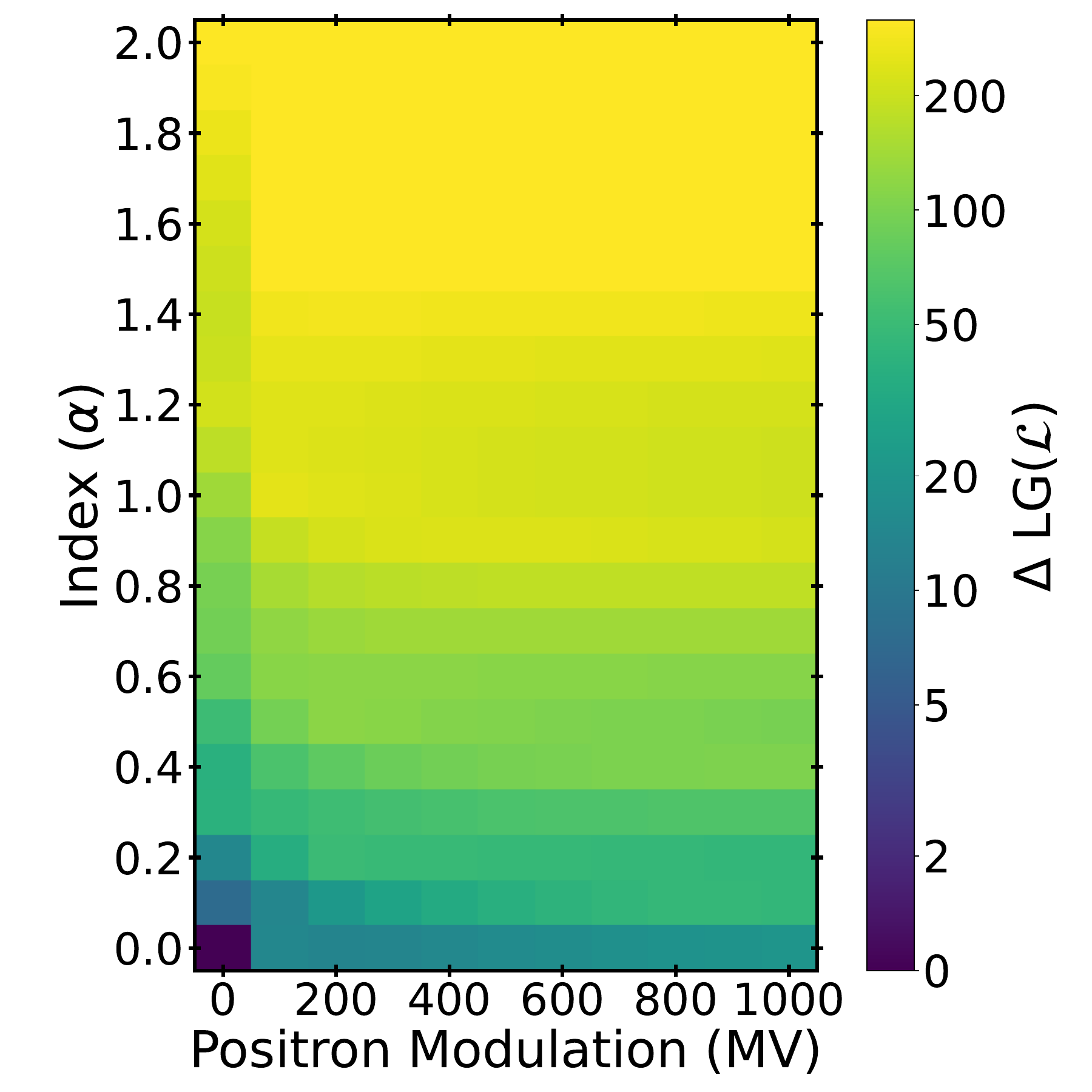}
\caption{The best fit solar modulation parameters following Model~III of our analysis, where the electron and positron modulation parameters are modified by an energy-dependent power law with an index $\alpha$, which increases the modulation of low-energy electrons. We find no preference for such a model, with a best-fit value and 5$\sigma$ error of $\alpha$~=~0.0$^{+0.17}$. We note that the value of $\alpha$ is restricted to be identical for electrons and positrons. The uncertainties on the electron and positron modulation values are similar, due to the fact that $\alpha$=0 is the default value for our solar modulation model.}
\label{fig:solarmodulation_alpha}
\end{figure*}

Figure~\ref{fig:solarmodulation_energybins} shows the best-fit modulation potential (using Model~I in our analysis) for electrons in different energy bins of our analysis, spanning from the lowest- to highest-energy data. We find results that are relatively consistent between each model, with slight changes between preferences for 400~MV and 500~MV modulation potentials in most energy bins. For $\gamma$-ray energies above 3~GeV (corresponding to high energy \epm~that should not be strongly modulated) we find a slight preference for negligible modulation. However, for these electron energies, the effect of modulation is negligible regardless of the modulation potential, meaning the change in the $\gamma$-ray data between our models is very small. Instead, the small preference for low-modulation potentials likelihood corresponds to small errors in the high-energy LIS spectrum, which would drive these models towards low values of the solar modulation parameter.

We note that a weakness of the above approach stems from the fact that we are testing the modulation of cosmic-ray electrons, but binning the results in terms of the lower $\gamma$-ray energy, which is affected by \epm~across a wide-array of energy scales. In order to account for this issue, we developed the \emph{ad hoc} solar modulation Model~III in Eq.~\eqref{eq:model_energy_dep}, where we add a power-law correction into the solar modulation potential that makes modulation larger at low energies, and smaller at high energies (roughly fitting the small preference we see in Figure~\ref{fig:solarmodulation_energybins}). We analyze the data, binning our result into bins of $\alpha$=0.1 over a range $0\leq\alpha\leq2$. Because this multiplies the necessary computational time by a factor of 21, we restrict electrons and positrons to have the same value of $\alpha$, though they can have different normalizations for their modulation potentials. We additionally reduce the binning of our electron and positron normalizations to bin sizes of 100~MV, rather than bin sizes of 25~MV, which we use throughout most of the text.

Figure~\ref{fig:solarmodulation_alpha} shows the results for the best-fit modulation parameters of both electrons (left) and positrons (right), finding them to be unchanged from their default values. We find that $\alpha$=0.0 (no energy-dependent change in the modulation potential) provides the best fit for our analysis, and we find the best fit point to have $\Phi_0$=500~MV for electrons and 0~MV for positrons, consistent with our previous results.


\subsection{Time Variability of Solar Modulation}
\label{subsec:timevariability}

An important property of solar modulation is that it is variable on all time scales, tracking the strength, speed, and turbulence of the solar wind. Studies of the solar wind (or studies where the solar wind is a key nuisance parameter) must carefully account for the exact observational time frame. 

The principal change in the time-dependence of solar modulation is a transition from a relatively small modulation potential at solar minimum, to a larger potential at solar maximum, when the strength and turbulence of the heliospheric magnetic field increase. There are also important charge sign-dependent effects based on the 22-year polarity cycle of the Sun. During periods of positive polarity, negatively charged particles must diffuse and drift across the heliospheric current sheet, while positive charged particles diffuse more efficiently from polar regions of the solar system~\cite{2013LRSP...10....3P}. During periods of negative polarity the opposite is true, and the positively charged particles diffuse against the heliospheric current sheet, while the negative particles diffuse more efficiently from polar regions. This means that the solar modulation potentials that we calculate for electrons and positrons should differ as a function of time. However, as shown in Figure~\ref{fig:solarmodulation_heatmap}, we note that the positron modulation potentials are relatively poorly constrained, making it difficult to accurately model their time dependence.

\begin{figure*}[t!]
\centering
\includegraphics[width=.98\textwidth]{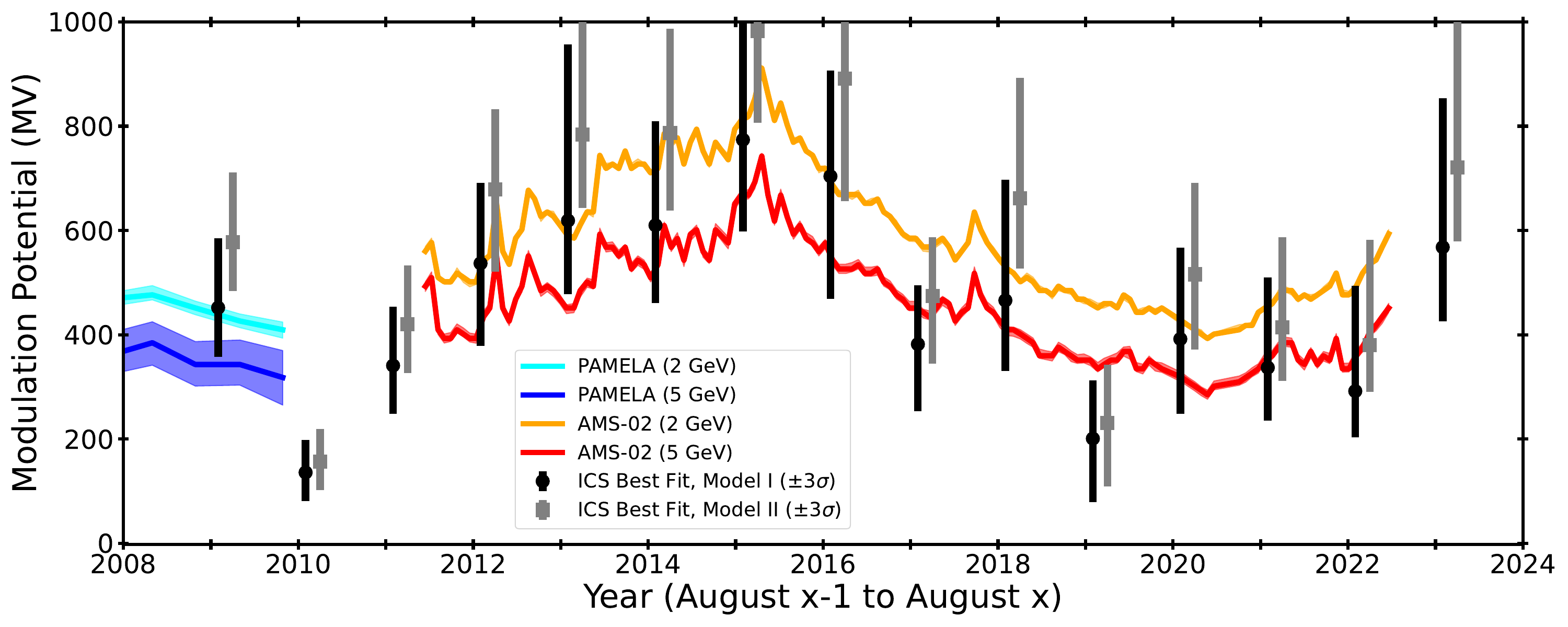}
\caption{The time dependence of the solar modulation potential at 1~AU for electrons. The best-fit modulation potential using Model I (black circle) in our analysis is calculated independently for data from each year and is shown along with 3$\sigma$ error bars surrounding the best-fit point, while the best-fit modulation potentials for Model II are shown in gray squares. Small offsets in the x-axis between the solar halo datasets are added to increase visibility, the dates of each analysis are identical. We find strong evidence for an increase in the solar modulation potential during the peak of solar activity near 2014, with lower potentials during solar minimum. The amplitude and time variation of solar modulation closely match PAMELA data taken from July 2006 and December 2009~\cite{Adriani:2015kxa}, and AMS-02 data taken from May 2011 to June 2022~\cite{PhysRevLett.134.051002}, calculated based on the electron flux observed at Earth. The systematic uncertainty in the solar modulation potential based on $\gamma$-ray observations (moving between Model I and Model II) is of a similar degree as the uncertainty in the modulation potential from cosmic-ray studies based on the minimum electron energy that is measured. }
\label{fig:timedependence}
\end{figure*}

Figure~\ref{fig:timedependence} shows the time-dependent modulation potential for cosmic-ray electrons in our solar halo models. This is calculated by splitting our data into one year time segments spanning between the August 4th periods of each year, and then calculating the best-fit log-likelihoods as a function of the solar modulation potentials for both electrons and positrons. We again minimize our fits to choose the best-fitting positron modulation potential for each value of the electron modulation potential. Our results are consistent with the theoretically-expected time-dependence for the solar modulation of electrons. We find the smallest values of solar modulation during the 2008 and 2019 solar minima, with larger modulation values during the 2014 solar maximum.

Figure~\ref{fig:timedependence} also compares our results against the solar modulation potential at 1~AU for electrons calculated using PAMELA and AMS-02 data. PAMELA has measured the time-dependent electron flux from July 2006 to December 2009 during the solar minimum~\cite{Adriani:2015kxa}, while AMS-02 provides data for each Bartels rotation from May 2011 to June 2022~\cite{PhysRevLett.134.051002}. For each given time data point, we calculate the best-fit solar modulation potential using the force-field solution (Eq.~\eqref{eq: force field potential}). We include only the electron data above either 2~GeV or 5~GeV, as electrons below these energy ranges would primarily produce $\gamma$-rays below our considered energy threshold. The results from PAMELA are given in cyan and blue and from AMS-02 in orange and red, for the $>$2~GeV and $>$5~GeV cases, respectively, including a 1$\sigma$ uncertainty band. We find a close match between the time variation in our observations of the solar modulation potential, and those from local electron measurements. We find that the level of systematic uncertainty that is introduced by utilizing Model I or Model II is similar to the degree of systematic uncertainty that stems from well-motivated choices of different minimum electron energies in force-field models of solar modulation based on local electron measurements. These results validate both the physical nature of the solar halo (as a result of the same electron population that is observed at the Earth), as well as the standard picture where the majority of solar modulation occurs between the edge of the heliosphere and 1 AU, with only small deviations from standard force-field potential models within the inner solar system. We note that these results also closely match previous results on the time variation of the electron modulation potential during solar cycle 24 by Ref.~\cite{Cholis:2022rwf}.

\subsection{Azimuthal Distribution}
\label{subsec:asphericity}

Finally, we note that models of solar modulation include both an \epm~population that travels slowly across the ecliptic plane, and a second population that travels to the earth primarily from polar regions above and below the plane. The relative electron and positron content of this flux changes when the solar polarity flips.

In addition to testing the radial morphology of the solar halo, we can also search for azimuthal asymmetries in the halo potential that may stem from this change in the \epm~flux along and away from the ecliptic plane. In this paper, we examine these asymmetries in the context of a force-field model based on Model I, but where we adopt different modulation potentials for emission along the $T_x$ plane, which lies along the equatorial plane, and the $T_y$ plane, which along its rotational axis. We calculate the solar halo profile at any point $\left( T_x, T_y \right)$ with a modified modulation potential we name $\Phi^a_0$ as:

\begin{multline}
    \Phi^a_{0}\left( T_x, T_y \right) = \sin^2( \tan^{-1}(T_x,T_y))\Phi_{0, Tx} \\ + \cos^2( \tan^{-1}(T_x,T_y))\Phi_{0, Ty}
\end{multline}

\noindent where $\Phi_{0, Tx}$ and $\Phi_{0, Ty}$ are flux maps that represent different choices of the modulation potential defined at 1~AU. We calculate one value of $\Phi^a_0$ for electrons, and a second $\Phi^a_0$ for positrons. This produces an equivalently normalized $\gamma$-ray flux map at each point in the sky, with an intensity that rotates with the azimuthal coordinate. We note that due to projection effects, the map is not exactly split along a 45$^\circ$ axis. However, there is no particular reason for the asphericity of the solar modulation signal to be a simple harmonic, and thus we do not need to be extremely specific about our choice. In this section we are only testing for the possibility of some non-symmetric profile in our results.

\begin{figure*}[t!]
\centering
\includegraphics[width=.48\textwidth]{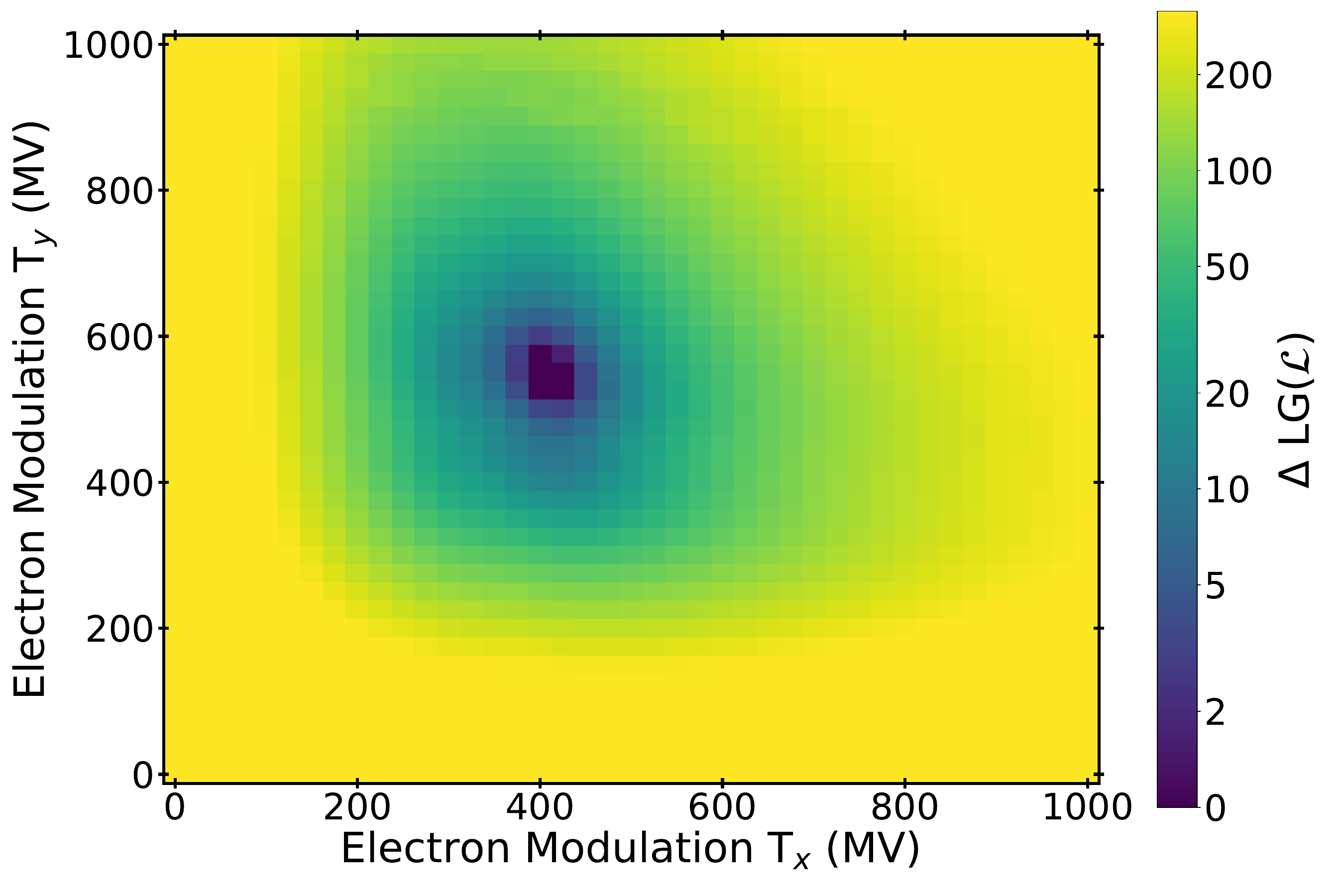}
\includegraphics[width=.48\textwidth]{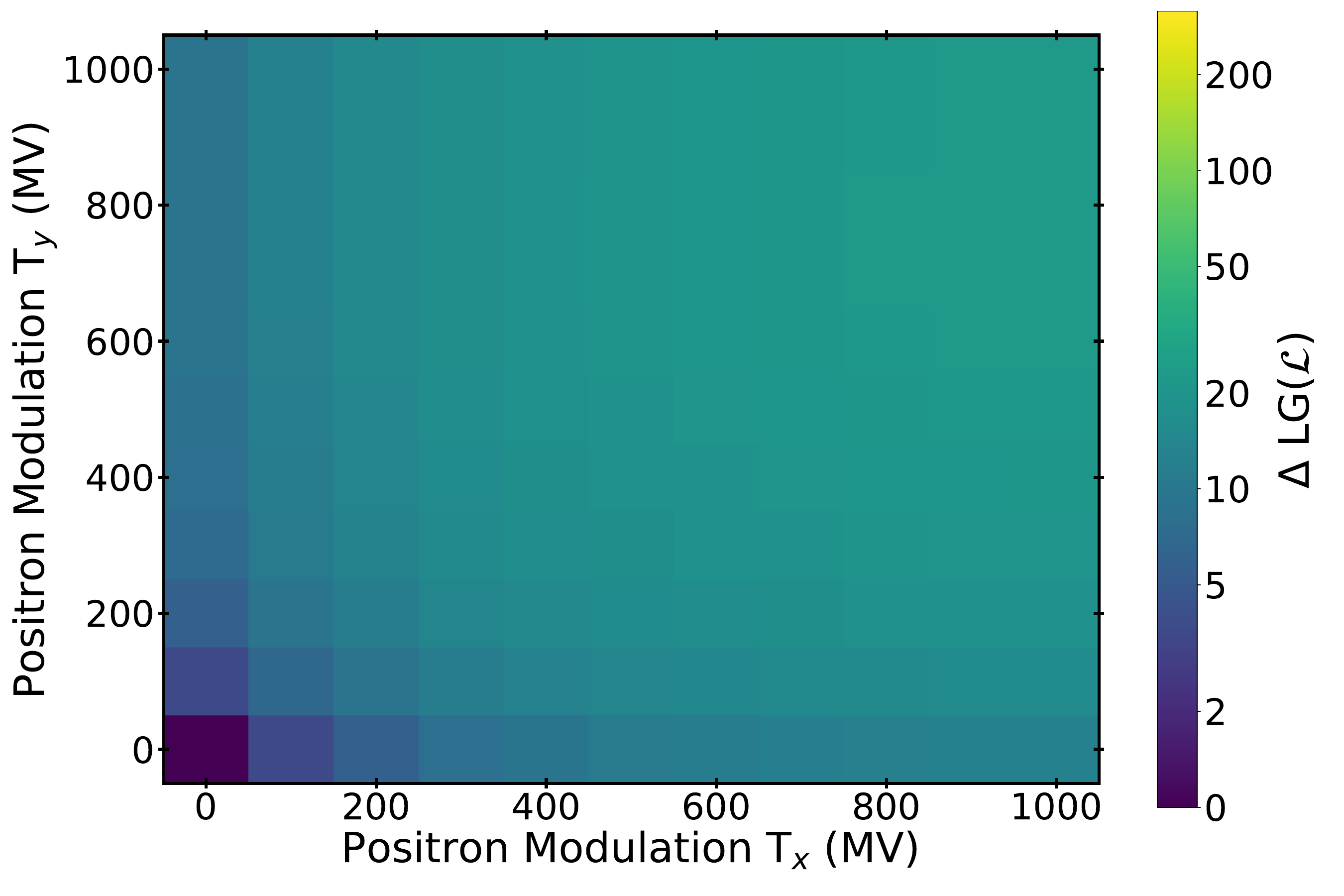}
\caption{The azimuthal distribution of the solar halo flux, which is calculated by assuming two different modulation amplitudes in the $T_x$ and $T_y$ directions for both electrons (left) and positrons (right). The log-likelihoods are obtained by minimizing the fits over leptons of the opposite charge. We find only a small evidence for an azimuthal asymmetry, which indicates excess emission along the ecliptic plane, with a log-likelihood preference of 11.0. However, it is difficult to determine whether this preference corresponds to a smaller modulation for electrons across the ecliptic plane, or the fact that our 15 year observation time includes a longer period when electrons traveled through the ecliptic plane.}
\label{fig:asymmetry}
\end{figure*}

Figure~\ref{fig:asymmetry} shows the result of this analysis. We again independently study modulation potentials for electrons and positrons by minimizing our fits over the other parameter. Due to the large number of choices of  modulation potentials for our analysis , we use a coarser grid of 100~MV for positron normalization potentials, which affect our results much less than the electron modulation potential. This leaves us with 41$^2 \times$11$^2$ simulations. Interestingly, our analysis finds some evidence for an azimuthal asymmetry in the electron spectrum. Our models prefer a modulation potential of 550~MV in directions perpendicular to the ecliptic plane, but a modulation potential of 400~MV in directions along the ecliptic plane. This result is preferred at a log-likelihood of 11.0, compared to the best fit symmetric result (with a modulation potential of 450~MV in both directions). This formally corresponds to a 4.7$\sigma$ detection of an asymmetry in the solar modulation potential, though we note that systematic errors in our analysis may make it difficult to conclusively argue that there is an overall asymmetry in the modulation potential.

However, we note that solar modulation is expected to be time dependent, and the polarity flip of the Sun occurred in $\sim$2014, a little less than halfway through the analysis window (and closer to halfway through the total exposure due to the decreased solar exposure after the 2018 failure of the solar panel motor). After this period, electrons drifted primarily across the heliospheric current sheet (in the T$_x$ direction). Thus it is reasonable to assume that the change in modulation potentials does not correspond to any intrinsic difference in electron propagation in each direction, only that the number of electrons moving across the heliospheric current sheet over our 15 year period is larger than in polar regions. Still, our results provide some evidence that the electron population is not spherically symmetric within 1~AU.

\begin{figure*}[t!]
\centering
\includegraphics[width=1.0\textwidth]{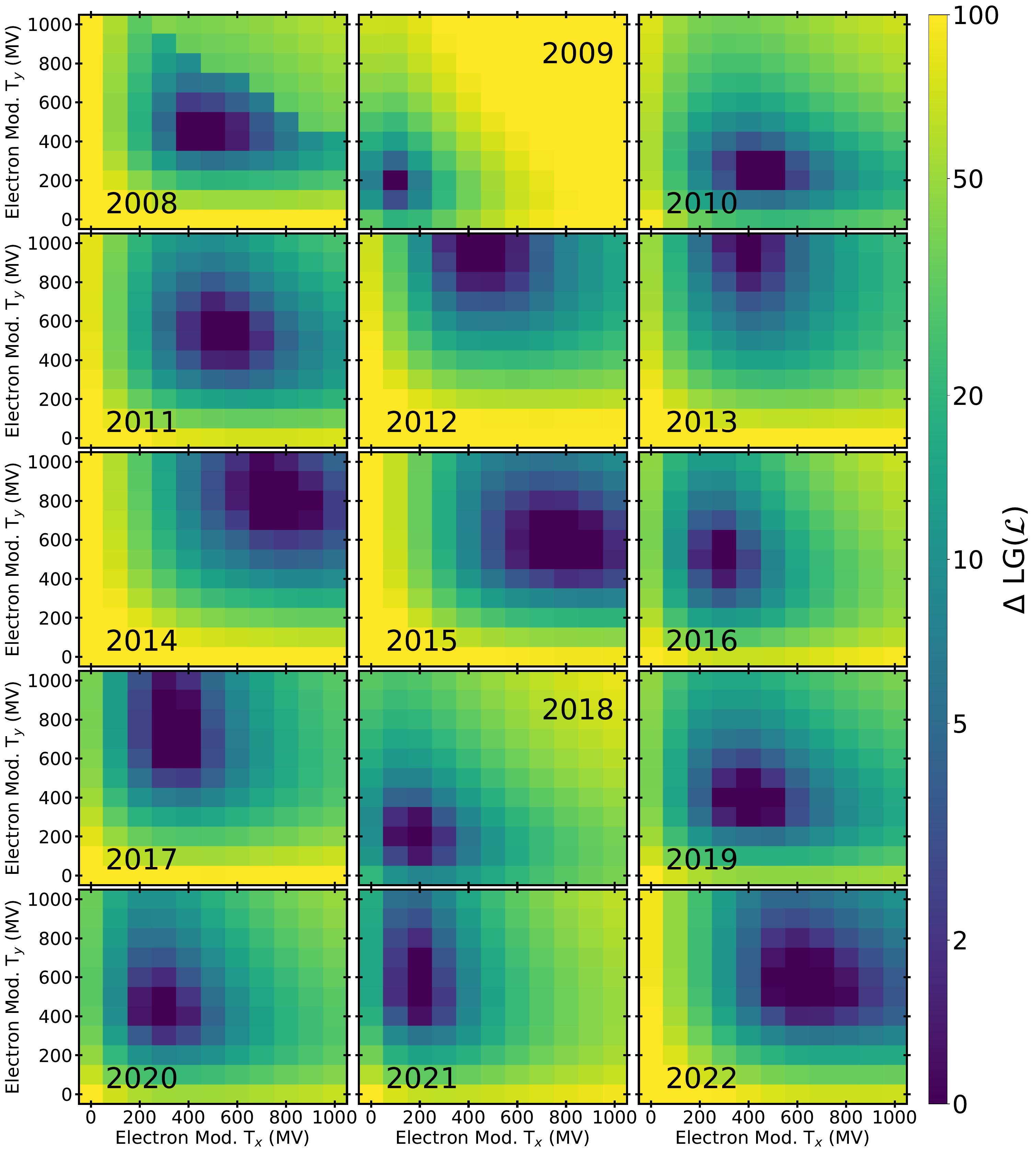}
\caption{Same as Figure~\ref{fig:asymmetry}, but showing the yearly asymmetry in the azimuthal distribution of the solar modulation. The time variation of the solar modulation potential is critical due both to the 12-year solar cycle and the $\sim$2014 polarity flip of the heliospheric magnetic field.  We only show results for electrons because the positron data is essentially unconstrained in small time windows. During the solar minimum periods (roughly 2008--2010 and 2018--2021) we find evidence for similar modulation potentials in the T$_x$ and T$_y$ directions. There are some hints of statistically significant asymmetries during solar maximum periods, but the trend does not closely follow the polarity of the heliospheric magnetic fields, meaning that temporal asymmetries may be dominated by small-scale effects. }
\vspace{1.0cm}
\label{fig:asymmetry_yearly}
\end{figure*}

In order to further investigate this result, we must probe the temporal change in the electron and positron modulation potentials. Figure~\ref{fig:asymmetry_yearly} shows the results for the yearly time variation of the electron modulation potential. We find that there is a pronounced change in the electron modulation in the T$_x$ and T$_y$ directions as a function of time. During the solar minimum periods (2008--2010 and 2017--2020), we tend to find fairly similar modulation potentials in the T$_x$ and T$_y$ directions. This is an expected result, as the small modulation potentials (and relatively non-turbulent nature of heliospheric magnetic fields) tend to minimize the azimuthal asymmetry of solar modulation.

Observations during solar maxima tend to have more pronounced differences between the ecliptic plane and polar regions of the solar system. During the first solar maximum (2012--2014), we tend to find larger modulation potentials in the $T_y$ direction. During the second maximum (primarily 2022), we find roughly equivalent modulation potentials in both directions. In some years (\emph{e.g.}, 2012, 2013, 2017, 2021) we do find significant asymmetry, verifying that these solar maximum periods can produce significant differences in solar modulation along and across the ecliptic plane. However, we do not see a clear pattern where the polarity of the magnetic field routinely makes emission brighter in the $T_x$ or $T_y$ directions. This potentially suggests that any asymmetry in electron propagation within the inner solar system may be dominated by smaller scale features that are not related to the overall polarity of the heliospheric magnetic field.


\section{Conclusions}
\label{sec:conclusions}

Observations of the solar halo provide a unique view into the effect of the heliosphere on cosmic-ray propagation. Importantly, most solar halo inputs, including the interstellar galactic cosmic-ray flux and the starlight density in the solar system, are extremely well known. This means that our observations directly constrain the heliospheric magnetic field and its effect on cosmic-ray propagation. Precise observations of cosmic rays in the heliosphere have critical implications for space physics, cosmic-ray physics, and dark matter searches. By searching for halo $\gamma$-rays, we can extend these observations into azimuthal and distance regimes that are not currently probed by direct cosmic-ray measurements. 

To precisely observe the solar halo, we developed a world-leading model capable of differentiating $\gamma$-ray emission that traces moving solar system objects from the remainder of the astrophysical background. These diffuse maps reach 1\% precision in fitting the $\gamma$-ray data near the Sun over most of our representative energy range, far exceeding the precision of theoretically motivated diffuse models of the Fermi-LAT sky. The accuracy of these models has allowed us to, for the first time, detect the solar halo in an energy range between 31.6~MeV and 100~GeV and out to 45$^\circ$ from the Sun. Moreover, the large dataset unlocked by our analysis allowed us to sensitively probe the time variation of solar $\gamma$-rays and search for azimuthal asymmetries in the solar halo, which could provide new insights into the interactions of cosmic rays with the solar wind and heliospheric current sheet.

Our observations are well fit by simple theoretical models of the solar halo, which are themselves well-constrained by cosmic-ray \epm~observations near Earth~\cite{Cholis:2015gna, Cholis:2020tpi, Cholis:2022rwf, Kuhlen:2019hqb}. Utilizing a simplistic force-field model for solar modulation, we find a best-fit modulation potential (normalized at 1~AU) for electrons of $\sim$500~MV. We find a smaller modulation potential for positrons, but note that there are large uncertainties on the positron modulation because: (1) it is subdominant, and the solar halo $\gamma$-ray emission is agnostic as to its electron or positron origin, (2) small errors in the LIS spectra will have a disproportionate effect on the smaller positron component. The large statistical dataset enabled by our astrophysical modeling allows us to sub-divide the data and obtain the first yearly measurement of solar modulation based on $\gamma$-ray data. Our best-fit force-field models closely match local observations of cosmic-ray electrons and positrons.

Critically, our $\gamma$-ray analysis allows us to perform tests of solar modulation that are not accessible to studies based on local cosmic-ray measurements. Notably, we can chart the radial distribution of the \epm~population in regions within 1~AU. We find: (1) no evidence for a decrease in the effect of solar modulation between the Earth and 0.1~AU, compared to standard force-field potentials (e.g., Ref.~\cite{Li:2022zio}), (2) no evidence for an additional energy-dependent variation in the solar modulation potential, aside from the energy dependence of a standard force-field model, (3) while we find evidence for yearly changes in the azimuthal symmetry of modulation, these do not appear to cleanly match theoretical expectations based on the polarity of the heliospheric magnetic field. Unlike the case of solar disk $\gamma$-rays, where the $\gamma$-ray data has numerous features that were not expected by theory~\cite{Tang:2018wqp, Linden:2018exo, Linden:2020lvz, Arsioli:2024scu}, the origin of the solar halo appears well-matched to our theoretical understanding of \epm~propagation in the inner heliosphere, unlocking new studies of both heliospheric physics, as well as beyond standard model searches for rare interactions in the inner solar system.

\bigskip
\textbf{\textit{Note added:}}
After this paper was submitted to arXiv, a related result by the Fermi-LAT collaboration appeared~\cite{Acharyya:2025xya}. Unlike our paper, this paper does not develop a full background model for astrophysical $\gamma$-rays, nor does it calculate the spectrum or radial distribution of the solar halo emission. However, it overlaps our paper in that it also focuses on the time variations of the solar disk and solar halo fluxes relative to their averages.

Their disk results are consistent the results we previously produced in Ref.~\cite{Linden:2020lvz}, but their work covers a longer time span. Their halo results can be compared to ours in Figure~\ref{fig:timedependence} here, though we calculate the true flux of the solar halo, allowing us to plot the results in terms of the solar modulation potential instead of the fractional flux variation.  In the time period 2008-2012, their results are qualitatively consistent with ours.  Starting in 2013, the solar halo flux decreases in our analysis (that is, the modulation potential increases), which is the expected result from both AMS-02 cosmic-ray data and also the variation of the solar disk flux.  However, the Fermi-LAT halo flux increases, which is surprising, as they note.  Further investigations are warranted (and ongoing), however, there are several critical details missing from the methodology of Ref.~[55] (for example how the exposure is calculated), and we have not been able to obtain information about how critical analysis steps were performed.  The sophisticated background modeling in our paper appears to be an important aspect of the differences between the results.

\medskip
\section*{Acknowledgments}
We would like to thank Dan Hooper, Zhe Li, Igor Moskalenko, and especially Ilias Cholis for helpful comments which improved the quality of this manuscript. T.L and M.C. are supported by the Swedish Research Council under contract 2022-04283 and the Swedish National Space Agency under contract 117/19. J.\,T.\,L. and A.\,H.\,G.\,P are supported by NASA Grant Nos.\ {80NSSC25K7761}, {80NSSC20K1354}, and {80NSSC22K0040}.
B.\,Z. is supported by Fermi Forward Discovery Group, LLC under Contract No. 89243024CSC000002 with the U.S. Department of Energy, Office of Science, Office of High Energy Physics.
I.\,J. acknowledges support from the Research grant TAsP (Theoretical Astroparticle Physics) funded by INFN, and Research grant ``Addressing systematic uncertainties in searches for dark matter'', Grant No.\ 2022F2843L, CUP D53D23002580006 funded by the Italian Ministry of University and Research (\textsc{mur}). 
J.\,F.\,B. is supported by the National Science Foundation Grant No.\ {PHY-2310018}.
Parts of this work were performed using computing resources provided by the National Academic Infrastructure for Supercomputing in Sweden~(NAISS) under Projects~\mbox{2023/3-21}, \mbox{2023/6-297}, \mbox{2024/5-666} and~\mbox{2024/6-339}, which is partially funded by the Swedish Research Council through Grant~\mbox{2022-06725}.


\bibliography{main}


\appendix

\section{High-Statistics Modeling}
\label{sec:highstatistics}

Throughout the paper we employ a ``low background" model, which cuts approximately 81\% of the total solar exposure from our study. In principle, such a cut should be unnecessary because our ``on/off" model should exactly calculate (to the level of statistical noise) the astrophysical $\gamma$-ray flux due to non-solar effects. This method has been utilized in Refs.~\cite{Linden:2020lvz, Leane:2021tjj} and has demonstrated significant robustness.

\begin{figure}[t!]
\centering
\includegraphics[width=.48\textwidth]{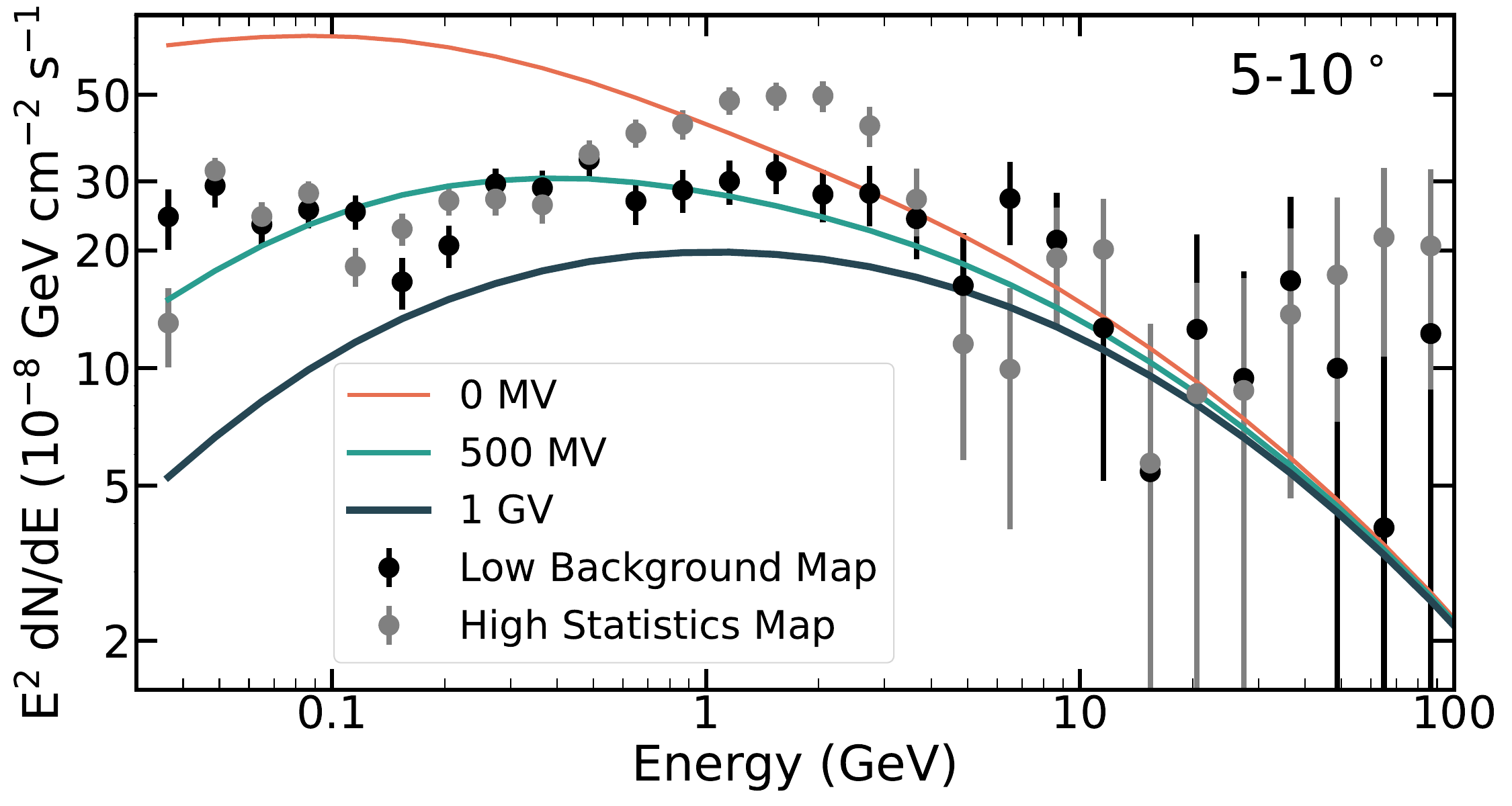}
\includegraphics[width=.48\textwidth]{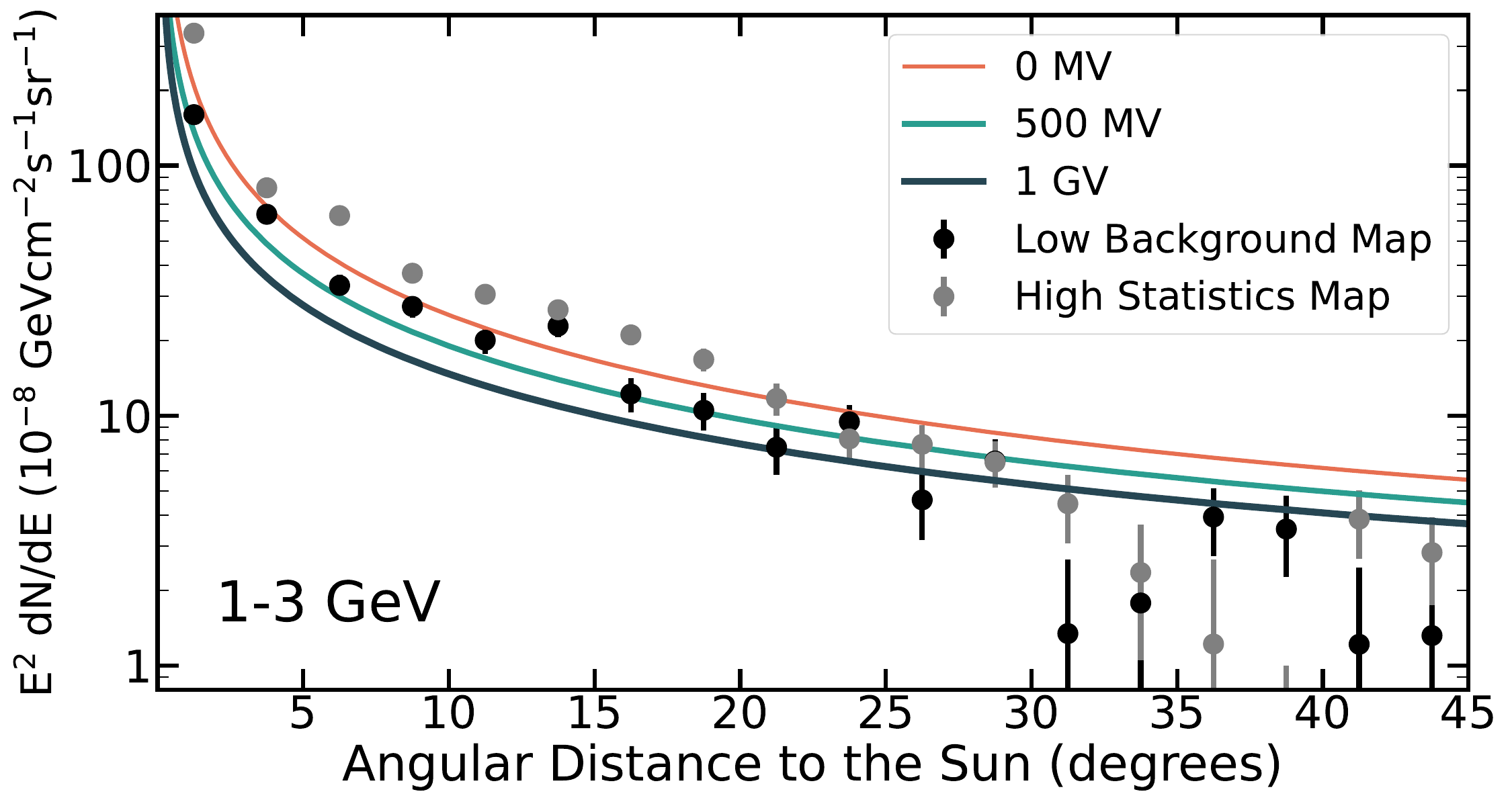}
\caption{Same as Fig.~\ref{fig:main}, but with the inclusion of the high-statistics model, which includes 100\% of the $\gamma$-ray data, including emission from the galactic plane, and emission from bright point sources. In general, the maps are in relative agreement, with the exception of a significant 1--3~GeV spectral bump that appears in the data at nearly all radial extensions. Notably, this spectral feature is similar to the 1--3~GeV excess first found in the data surrounding solar events in Refs.~\cite{Tang:2018wqp, 2018arXiv181011394B}, but removed or mitigated in the {\tt p305} dataset.}
\label{fig:app:highstatisticscompare}
\end{figure}

In Figures~\ref{fig:app:highstatisticscompare}, we show the resulting solar halo ICS flux in our optimal angular and energy cuts, identical to the result shown in Figure~\ref{fig:main}. In general, we find a very close match between the luminosity of the solar halo between our models. However, in the top panel we observe a significant bump in the energy range between 1 and 3~GeV in our ``high-statistics map", which disappears in the ``low-background map". Focusing on this energy range (bottom), we see that this bump is most notably within $\sim$20~degrees of the solar position, and decreases at larger distances from the Sun. This indicates that it is not a simple mismodeling of the astrophysical background -- which would have a similar surface brightness throughout our ROI.

The odd 1--3~GeV spectral feature echoes the findings of Ref.~\cite{Tang:2018wqp}, where it was first noted that there was a statistically significant residual at energies between 1  and 3~GeV in solar observations. Discussions with members of the {\it Fermi}-LAT collaboration (for more details see discussions in Ref.~\cite{Tang:2018wqp, Linden:2020lvz}) determined that this residual was already known in Monte Carlo data, but was not expected to be observable in astrophysical data. The error stemmed from charged cosmic-ray leakage through ribbons in the anti-coincidence shield. Because these ribbons are oriented along the $\phi=0$ instrumental coordinate of the {\it Fermi}-LAT, this leakage is particularly bright near the Sun, due to the fact that the solar panels are also aligned with $\phi=0$. The p305 {\it Fermi}-LAT dataset, released in late 2018, was designed in part to eliminate this systematic error\footnote{https://fermi.gsfc.nasa.gov/science/mtgs/symposia/2018/program/89/}.

\begin{figure}[t!]
\centering
\includegraphics[width=.48\textwidth]{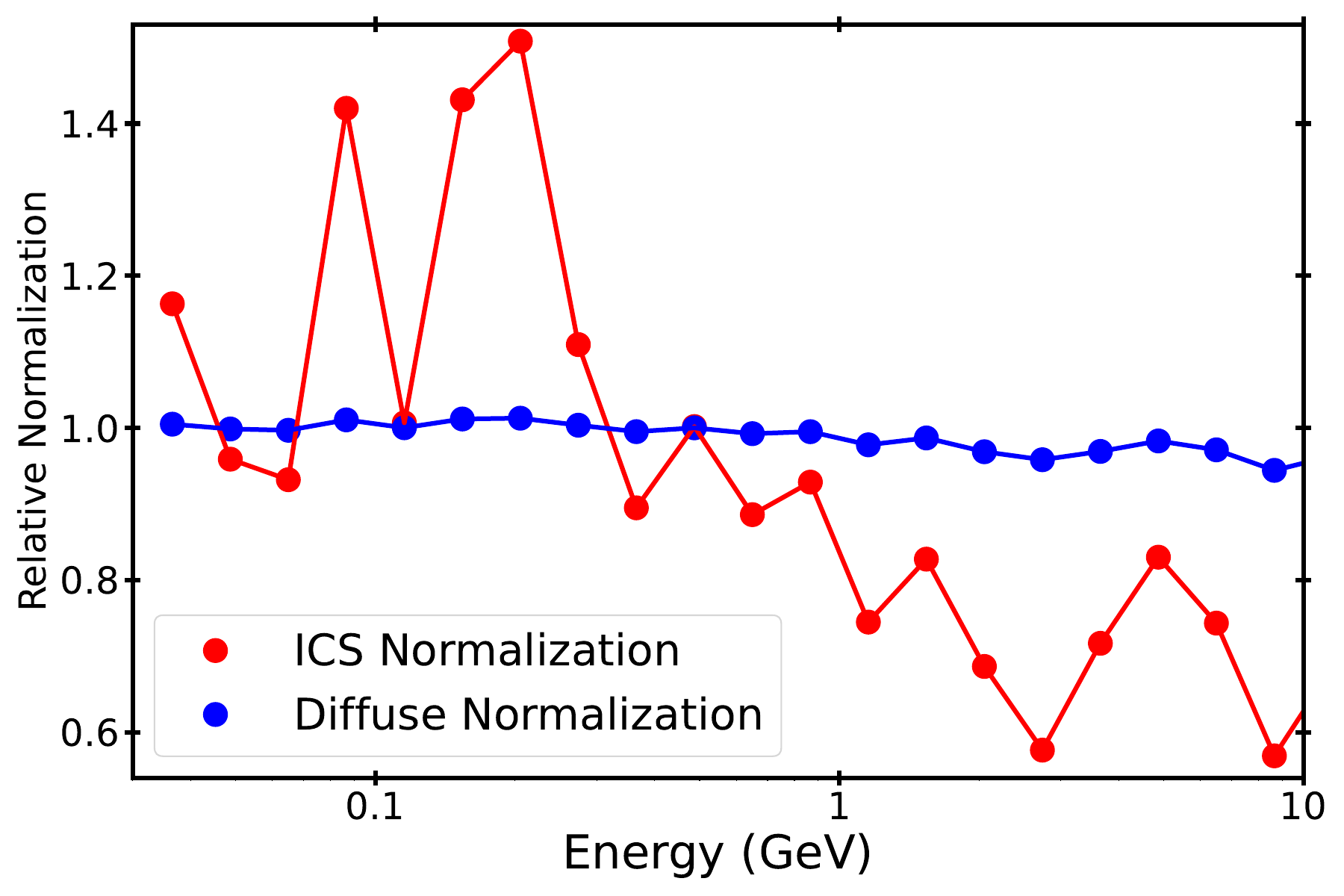}
\caption{The impact of allowing the diffuse normalization to float (from its default value of 1) independently in each energy bin of our analysis. We find that the model prefers diffuse normalizations within 4\% of the predicted value throughout the majority of our analysis energy range (and values within 1--2\% of the default value in most energy bins. However, this still has a significant impact on our ICS emission modeling, affecting the total ICS flux by a factor of $\sim$20\% in most energy bins, but by up to a factor of 2.}
\label{fig:floating_diffuse_normalization}
\end{figure}

Compared to solar disk observations, however, the solar halo has a surface brightness that is thousands of times smaller, meaning that any residual contamination in the diffuse emission may have large effects on the solar halo spectrum. In Figure~\ref{fig:floating_diffuse_normalization} we show that $\sim$1\% changes in the diffuse normalization can produce 10-20\% differences in the halo normalization, due to its much lower flux. We note that allowing the diffuse model to float in each of the 28 tested energy bins improves the log-likelihood of the fit by 128~, indicating that there is a statistical preference for such a modification, but it is relatively small compared to the preference of the ICS halo.

However, there is a problem with this simple picture. Our analysis finds that the background mismodeling rate is highest along the galactic plane -- which is bright due to $\gamma$-ray (and not cosmic-ray) emission. Cosmic-ray electrons are expected to be mostly isotropic. This opens a second possibility that there are systematic errors in the $\gamma$-ray effective area of the instrument as a function of the $\phi$-coordinate of the {\it Fermi}-LAT. This would cause the calculated astrophysical $\gamma$-ray background rate to be higher or lower at a rate that correlates with the $\gamma$-ray flux. 

The most effective way of testing such a scenario is to build a model of the {\it Fermi}-LAT effective area that isolates specific ranges of the $\phi$-coordinate and then calculates the resulting flux of the $\gamma$-ray sky. However, such a study is not currently possible within the standard version of the {\it Fermi}-LAT tools, which make it impossible to directly cut on the $\phi$-dependence of the {\it Fermi}-LAT effective area. Further testing is necessary to study this effect, in order to determine whether it is responsible for the 1--3~GeV bump in the $\gamma$-ray background.

We note that additional cuts (e.g., larger latitude cuts or larger point source masks) had a negligible effect on the resulting ICS template, and we consider the current masks to be sufficient for calculating the true flux of the solar halo.

\section{Solar Halo Skymaps}

\begin{figure*}[t!]
\centering
\includegraphics[width=.99\textwidth]{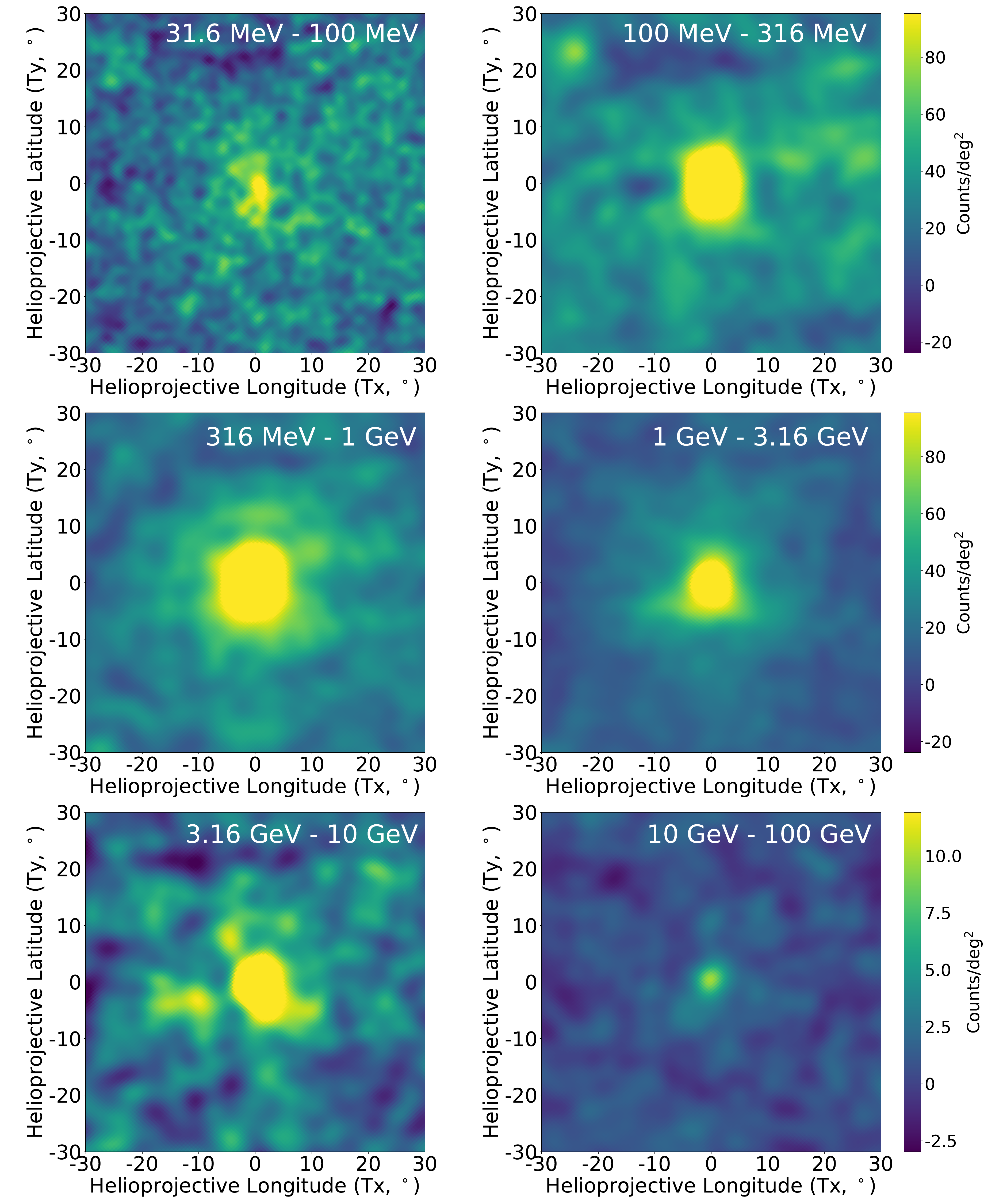}
\caption{The reconstructed solar halo emission in six-representative energy ranges: 31.6--100~MeV, 100--316~MeV, 0.316--1~GeV, 1--3.16~GeV, 3.16--10~GeV, and 10--100~GeV. The plots show a combination of the solar halo emission plus all residual emission that is not absorbed by either the diffuse background or solar disk template. The halo is robustly detected in all energy bins.}
\label{fig:halo_skymaps}
\end{figure*}

In Figure~\ref{fig:halo_skymaps} we show illustrated skymaps of the solar halo in six energy ranges that span our analysis: 31.6~MeV -- 100~MeV, 100~MeV -- 316~MeV, 316~MeV -- 1~GeV, 1~GeV -- 3.16~GeV, 3.16~GeV -- 10~GeV and finally 10~GeV -- 100~GeV. We stress that our quantitative analysis is divided into eight logarithmic energy bins per decade spanning the range from 31.6~MeV to 100~GeV, and the results shown here are produced by summing the results from several energy bins. These results are shown using the ``high-statistics" model, which has more uniform exposure over the sky and thus better illustrates the morphology of the solar halo emission.

Our results show the combination of all emission that is not absorbed by either the astrophysical background model or the solar disk template. Thus, it includes both the solar halo as well as any residual emission that is not accounted for in our model. For this plot, however, we smear the resulting map with a 4$^\circ$ gaussian PSF in order to better highlight large scale features such as the halo itself. 

Our results show that the solar halo is robustly detected and roughly azimuthally symmetric in every energy range in our analysis -- and is by far the brightest residual feature in our maps.



\end{document}